\documentclass[12pt,twoside,a4paper]{article}
\usepackage{amsmath,amssymb,latexsym,theorem,natbib,epsfig,color,subfigure,footmisc,bbm}
\usepackage{multirow,graphicx,array,rotating,url}  
\usepackage{epstopdf,afterpage,wasysym}
\usepackage{verbatim}
\usepackage[normalem]{ulem}
\def\crps{\mathop{\hbox{\rm CRPS}}}
\def\es{\mathop{\hbox{\rm ES}}}
\def\ess{\mathop{\hbox{\rm ESS}}}
\def\vs{{\rm VS}}

\def\rank{\mathop{\hbox{\rm rank}}}
\def\ri{\mathop{\hbox{\rm RI}}}

\def\md{\mathrm{MD}}

\numberwithin{equation}{section}

\title{Comparison of multivariate post-processing methods using global ECMWF ensemble forecasts}

\author{{M\'aria Lakatos}$^{1,2}$, {Sebastian Lerch}$^3$, {Stephan Hemri}$^4$ and {S\'andor Baran}$^{1,*}$ \vspace*{0.5cm}\\
{\small $^1$Faculty of Informatics, University of Debrecen, Hungary}\\
{\small $^2$Doctoral School of Informatics, University of Debrecen, Hungary}\\
{\small $^3$Institute of Economics, Karlsruhe Institute of Technology, Germany}\\
{\small $^4$Department of Mathematics, University of Zurich, Switzerland}
}  

\date{}

\begin{document}

\maketitle

\footnotetext[1]{Corresponding author: \url{baran.sandor@inf.unideb.hu}}.
\begin{abstract}
An influential step in weather forecasting was the introduction of ensemble forecasts in operational use due to their capability to account for the uncertainties in the future state of the atmosphere. However, ensemble weather forecasts are often underdispersive and might also contain bias, which calls for some form of post-processing. A popular approach to calibration is the ensemble model output statistics (EMOS) approach resulting in a full predictive distribution for a given weather variable. However, this form of univariate post-processing may ignore the prevailing spatial and/or temporal correlation structures among different dimensions. Since many applications call for spatially and/or temporally coherent forecasts, multivariate post-processing aims to capture these possibly lost dependencies.

We compare the forecast skill of different nonparametric multivariate approaches to modeling temporal dependence of ensemble weather forecasts with different forecast horizons. The focus is on two-step methods, where after univariate post-processing, the EMOS predictive distributions corresponding to different forecast horizons are combined to a multivariate calibrated prediction using an empirical copula. Based on global ensemble predictions of temperature, wind speed and precipitation accumulation of the European Centre for Medium-Range Weather Forecasts from January 2002 to March 2014, we investigate the forecast skill of different versions of Ensemble Copula Coupling (ECC) and Schaake Shuffle (SSh). In general, compared with the raw and independently calibrated forecasts, multivariate post-processing substantially improves the forecast skill. While even the simplest ECC approach with low computational cost provides a powerful benchmark method, recently proposed advanced extensions of the ECC and the SSh are found to not provide any significant improvements over their basic counterparts.

\end{abstract}

\section{Introduction}
\label{sec1}

In many cases, ensemble forecasts generated by numerical weather prediction (NWP) models suffer from systematic biases and underdispersion. In order to correct for these issues, over the last two decades a plethora of different post-processing approaches have been developed. They range from classical regression-type approaches like ensemble model output statistics \citep[EMOS;][]{grwg05} or Bayesian model averaging \citep[BMA;][]{rgbp05} to sophisticated machine learning based approaches as, for instance, applied in \citet{sl22}. \citet{vbd21} provide a comprehensive summary of statistical post-processing.

Most post-processing approaches implicitly assume statistical independence between different forecast margins like lead times, locations or weather variables. Apparently, this assumption does not allow for providing realistic forecast scenarios. However, end-users may be interested in scenarios like total precipitation over an entire hydrological catchment, the temporal evolution of precipitation, or the interaction of precipitation and temperature, in particular when temperature is close to zero degree Celsius. Therefore, for many use cases dependence structures need to be re-established explicitly in a second post-processing step after univariate calibration. To this end, different copula based approaches have been proposed. Most state-of-the-art multivariate post-processing applications employ an empirical copula based on a dependence template stemming either from a NWP ensemble or historical observations. The former and the latter are referred to as ensemble copula coupling \citep[ECC;][]{stg13} and Schaake shuffle \citep[SSh;][]{cghrw04}, respectively. We refer to \citet{sm18} for a detailed discussion of copula based methods to incorporate dependence structures.

\citet{lbm20} performed an extensive simulation study comparing several variants of ECC with SSh and the parametric Gaussian copula approach \citep[GCA;][]{mlt13}, and the authors concluded that the benchmark methods ECC-Q, that is ECC with equidistant quantile sampling \citep{stg13}, and SSh generally performed well. Moreover, their results suggest that more sophisticated approaches like, for instance, dual ECC \citep[dECC;][]{bbhtp16} or GCA outperform only in very specific conditions. In this study, we assess if these findings can be confirmed by using real NWP ensemble forecasts and observations. To this end, we apply univariate EMOS combined with a whole range of ECC and SSh variants to NWP ensemble forecasts of temperature, wind speed, and precipitation provided by the European Centre for Medium-Range Weather Forecasts (ECMWF). Observations from SYNOP stations are used for verification. To the best of our knowledge, our work is the first to compare a large variety of state-of-the-art two-step methods for multivariate post-processing including recently proposed similarity-based Schaake shuffle approaches and dual ECC.

Since we focus on the multivariate post-processing step, for the univariate calibration we follow simply the EMOS implementations by \citet{hspbh14}. However, for the multivariate step we compare three naive approaches, which are derived more or less directly from univariate EMOS, with four variants of ECC and five variants of the SSh. For the sake of simplicity and comprehensibility, we focus on temporal dependence.

After a short description of the data in Section \ref{sec2}, we provide a detailed summary of both univariate EMOS and all approaches we apply to re-establish the multivariate dependence structure in Section \ref{sec3}. In Section \ref{sec4} we present our results, followed by a brief discussion and conclusions in Section \ref{sec5}.

\section{Data}
\label{sec2}
Comparison of the various multivariate post-processing methods is performed with the help of global ECMWF ensemble forecasts of 2m temperature (T2M), 10m wind speed (V10) and 24h precipitation accumulation (PPT24) for the period between 1 January 2002 and 20 March 2014. The datasets at hand are identical to the ones investigated in \citet{hspbh14} containing the ECMWF high-resolution (HRES) forecasts, the 50-member operational ensemble (ENS) generated using random perturbations and the control run (CTRL) initialized at 1200 UTC with forecast horizons ranging from 1 day to 10 days, together with the corresponding observations. After an initial quality control removing SYNOP stations with missing or irregular data, 4160, 4388 and 2917 stations covering the entire globe remained for T2M, V10 and PPT24, respectively. For more details about the investigated data and the applied quality control procedure see \citet{hspbh14}.

\section{Methods}
\label{sec3}

As mentioned in the Introduction, we restrict our attention to two-step approaches to multivariate post-processing, where after an initial univariate calibration, multivariate predictions are obtained by combining the individual post-processed forecasts with the help of an empirical copula.

In what follows, let \ $\boldsymbol f^{(\ell)} = \big(f_1^{(\ell)},f_2^{(\ell)}, \ldots ,f_{52}^{(\ell)}\big)$ \ denote a 52-member ECMWF ensemble forecast with a lead time of \ $\ell$ \ days \ ($\ell =1,2, \ldots ,10$) \ initialized at a given time point, where \ $f_1^{(\ell)}=f_{\text{HRES}}^{(\ell)}$ \ and  \ $f_2^{(\ell)}=f_{\text{CTRL}}^{(\ell)}$ \ are the high-resolution and the control member, respectively, whereas \ $f_3^{(\ell)},f_4^{(\ell)}, \ldots ,f_{52}^{(\ell)}$ \ correspond to the 50 statistically indistinguishable (and thus exchangeable) ensemble members \ $f_{\text{ENS},1}^{(\ell)},f_{\text{ENS},2}^{(\ell)}, \ldots ,f_{\text{ENS},50}^{(\ell)}$ \ generated using random perturbations.

\subsection{Univariate post-processing}
\label{subs3.1}

For calibrating ensemble forecasts for a given location and time point with a given forecasts horizon, one can choose from a multitude of state-of-the-art approaches, as mentioned in the Introduction. Here we consider the computationally simple and efficient EMOS method, where post-processed forecasts are obtained in the form of parametric predictive distributions with parameters depending on the corresponding ensemble predictions. EMOS models for various weather quantities differ in the parametric distribution family to be used and in link functions relating the parameters of the predictive distribution to the raw ensemble forecasts. For univariate calibration of T2M, V10 and PPT24 ensemble forecasts, we make use of the specific EMOS approaches of \citet{hspbh14} described briefly in Sections \ref{subs3.1.1}, \ref{subs3.1.2} and \ref{subs3.1.3} below, respectively. To simplify notation we omit the indication of the forecast horizon and use notation \ $f_k=f_k^{(\ell)}, \ k=1,2, \ldots ,52,\ $ in these sections.

\subsubsection{Temperature}
\label{subs3.1.1}
The normal distribution and its generalizations (skewed normal, mixture of normals, etc.) are widely used to model temperature \citep[see e.g.][]{grwg05,rgbp05,rl18,t21}. The EMOS predictive distribution suggested by \citet{gneiting14} specifically for the 52-member T2M ECMWF ensemble is Gaussian with mean \ $\mu$ \ and variance \ $\sigma^2$ \ given as
\begin{equation*}
  \mu = a_0+a_1^2f_{\text{HRES}}+a_2^2f_{\text{CTRL}}+a_3^2\overline f_{\text{ENS}} \qquad \text{and} \qquad \sigma^2=b_0^2+b_1^2S_{\text{ENS}}^2,
\end{equation*}
where \ $\overline f_{\text{ENS}}$ \ and \ $S_{\text{ENS}}^2$ \ denote the mean and the variance of the 50 exchangeable ensemble members, respectively, that is
\begin{equation*}
  \overline f_{\text{ENS}}:=\frac 1{50}\sum_{k=1}^{50}f_{\text{ENS},k} \qquad \text{and} \qquad S_{\text{ENS}}^2:=\frac 1{50}\sum_{k=1}^{50}\big(f_{\text{ENS},k} - \overline f_{\text{ENS}}\big)^2.
\end{equation*}
Model parameters \ $a_0 , a_1, a_2, a_3, b_0, b_1 \in{\mathbb R}$ \ are estimated according to the optimum score principle of \citet{gr07}, that is by optimizing the mean value of an appropriate verification metric (see Section \ref{subs3.3}) over the training data consisting of past forecast-observation pairs.

To account for seasonal variations in temperature, inspired by \citet{sb14}, \citet{hspbh14} suggest a more complex model for the mean of the predictive distribution 
in the form of
\begin{equation}
  \label{regT2M}
  y_t=c_0+c_1\sin \left(\frac{2\pi t}{365}\right)+c_2\cos \left(\frac{2\pi t}{365}\right) +\varepsilon_t, \qquad t\in 1,2,\ldots ,n,
\end{equation}
where the dependent variables \ $y_t$ \ are either temperature observations for a given location or functionals of the corresponding ECMWF  ensemble forecast with a given lead time \ $\ell$, \ namely the high-resolution member \ $f_{\text{HRES}}$, \ the control run \ $f_{\text{CTRL}}$ \ and the mean of the exchangeable forecasts \ $\overline f_{\text{ENS}}$ \ from a training period of length \ $n$. \ With the help of model \eqref{regT2M} one can calculate the \ $\ell$-step ahead predictions \ $\widehat y$ \ and \ $\widehat f_{\text{HRES}}, \widehat f_{\text{CTRL}}, \widehat f_{\text{ENS}} $ \  of the observations and the corresponding functionals of the ensemble, respectively, and obtain a Gaussian predictive distribution with parameters
\begin{equation}
  \label{emosT2M}
  \mu = \widehat y +a_1^2 \big(f_{\text{HRES}}-\widehat f_{\text{HRES}}\big)+a_2^2\big(f_{\text{CTRL}}-\widehat f_{\text{CTRL}}\big)+a_3^2\big(\overline f_{\text{ENS}}- \widehat f_{\text{ENS}}\big) \quad \text{and} \quad \sigma^2=b_0^2+b_1^2S^2,
\end{equation}
where 
\begin{equation*}
   S^2:=\frac 1{52}\sum_{k=1}^{52}\big(f_k - \overline f\big)^2 \qquad \text{with} \qquad \overline f:=\frac 1{52}\sum_{k=1}^{52}f_k. 
\end{equation*}

\subsubsection{Wind speed}
\label{subs3.1.2}
To model wind speed one requires non-negative and skewed distributions. Here, we consider a normal law  \ $\mathcal N_0^{\infty}\big(\mu,\sigma ^2\big)$ \  with location \ $\mu$ \ and scale \ $\sigma>0$, \ left-truncated at zero, which is applied in the EMOS model of \citet{tg10}. As a natural transformation \citep[see e.g.][]{hr89}, \citet{hspbh14} model the square root of V10 as
\begin{equation}
  \label{emosV10}
  \mathcal N_0^{\infty}\Big(a_0+a_1^2\sqrt{f_{\text{HRES}}}+a_2^2\sqrt{f_{\text{CTRL}}}
  +a_3^2\sqrt{\overline f_{\text{ENS}}},b_0^2+b_1^2\md_{\sqrt f}\Big),
\end{equation}
where
\begin{equation*}
  \md_{\sqrt f}:=\frac 1{52^2}\sum_{k,\ell=1}^{52}\big|\sqrt{f_k}-\sqrt{f_{\ell}}\big|.
  \end{equation*}
The model parameters  \ $a_0 , a_1, a_2, a_3, b_0, b_1 \in{\mathbb R}$, \ can again be estimated using the optimum score estimation principle.

\subsubsection{Precipitation}
\label{subs3.1.3}
Statistical calibration of precipitation forecasts is more challenging than the post-processing of temperature or wind speed due to the discrete-continuous nature with a positive probability of observing zero precipitation. Following \citet{hspbh14}, to model PPT24 we consider the EMOS model of \citet{sch14}, where the predictive distribution is a generalized extreme value distribution left-censored at zero. Location \ $\mu$ \ and scale \ $\sigma$ \ are linked to the raw forecasts via
\begin{equation}
  \label{emosPPT24}
   \mu = a_0+a_1^2f_{\text{HRES}}+a_2^2f_{\text{CTRL}}+a_3^2\overline f_{\text{ENS}}+a_4^2\pi_0 \qquad \text{and} \qquad \sigma^2=b_0^2+b_1^2\md_f,
 \end{equation}
 where \  $a_0 , a_1, a_2, a_3, a_4, b_0, b_1 \in{\mathbb R}$, \
 \begin{equation*}
\pi_0:=\frac 1{52} \sum_{k=1}^{52}{\mathbbm{I}}_{\{f_k=0\}}  \qquad \text{and} \qquad \md_f:=\frac 1{52^2}\sum_{k,\ell=1}^{52}\big|{f_k}-{f_{\ell}}\big|,
\end{equation*}
are the proportion of ensemble members predicting zero precipitation and the ensemble mean difference, respectively, whereas shape \ $\xi$ \ is kept fixed \ ($\xi=0.2$).

\subsection{Multivariate methods}
\label{subs3.2}

Independent univariate calibration of ensemble forecasts \ $\boldsymbol f^{(\ell)}$ \ for each forecast horizon \ $\ell$ \ does not take into account the temporal dependencies between predictions initialized at the same time point. These dependencies are restored in a second step with the help of the approaches described below, which are based on empirical copulas.

In general, a copula is a multivariate cumulative distribution function (CDF) with standard uniform marginals. According to Sklar's theorem \citep{s59}, any $L$-dimensional CDF \ $H$ \ with marginal CDF-s \ $F^{(\ell)}, \ \ell =1,2, \ldots ,L$, \ can be decomposed as
\begin{equation*}
 H(x_1,x_2, \ldots ,x_L)=C\big(F^{(1)}(x_1),F^{(2)}(x_2), \ldots ,F^{(L)}(x_L)\big ), \qquad x_1,x_2, \ldots ,x_L \in{\mathbb R},
\end{equation*}
where \ $C$ \ is an $L$-dimensional copula representing the dependencies between the marginals. In our case \ $F^{(\ell)}$ \ is the predictive distribution corresponding to lead time \ $\ell = 1,...,10$ \ obtained after univariate post-processing, and we would like to combine calibrated predictions initialized at the same time point into a temporally consistent calibrated forecast trajectory represented by a 10-dimensional predictive CDF \ $H$.

Here we focus on non-parametric methods where marginals are represented by empirical CDFs \ $\widehat F^{(\ell)}$ \ of samples drawn independently from the corresponding predictive distributions, and \ $C$ \ is an empirical copula \citep{r09} providing a ``dependence template'' derived from a given discrete dataset. The learned dependence structure is then imposed on the post-processed forecasts by rearranging the marginal samples according to the specified template \citep[see e.g.][Section 8.4.2]{w19}.

\subsubsection{Ensemble copula coupling (ECC)}
\label{subs3.2.1}
In the ensemble copula coupling (ECC) approach the dependence template is obtained from the corresponding raw ensemble forecasts. The method, introduced by \citet{stg13}, consists of the following two simple steps.
\begin{enumerate}
\item For each dimension \ $\ell = 1,2, \ldots, L$, \ generate a sample \ $\widehat f_1^{(\ell)}, \widehat f_2^{(\ell)},  \ldots ,\widehat f_K^{(\ell)}$ \ of the same size as the raw ensemble (here \ $K=52$) \ from the calibrated marginal predictive distribution \ $F^{(\ell)}$, \ which  is assumed to be arranged in ascending order.
\item Consider permutations \ $\boldsymbol{\pi}_{\ell}=\big (\pi_{\ell}(1),\pi_{\ell}(2),\ldots ,\pi_{\ell}(K)\big)$ \ of \ $\{1,2,\ldots , K\}$ \ induced by the rank order structure of the raw ensemble \  $f_1^{(\ell)}, f_2^{(\ell)},  \ldots, f_K^{(\ell)}$, \ that is \ $\pi_{\ell}(k):=\rank \big(f_k^{(\ell)}\big)$ \ with ties resolved at random. The ECC calibrated sample \ $\widetilde f_1^{(\ell)}, \widetilde f_2^{(\ell)},  \ldots ,\widetilde f_K^{(\ell)}$  \ for dimension \ $\ell$ \ is obtained by rearranging the sample generated in step 1 according to permutation  \ $\boldsymbol{\pi}_{\ell}$, \ that is \
  \begin{equation*}
    \widetilde f_k^{(\ell)}:=\widehat f_{\pi_{\ell}(k)}^{(\ell)}, \qquad k=1,2, \ldots, K, \quad \ell=1,2,\ldots, L.
  \end{equation*}
\end{enumerate}  

Similar to \citet{lbm20}, we consider three different ECC variants depending on the sampling method in step 1. {\bf ECC-R} refers to random sampling from the predictive distribution \ $F^{(\ell)}$ \ and arranging the sample in ascending order, whereas in {\bf ECC-Q} one considers equidistant quantiles of  \ $F^{(\ell)}$, \ that is
\begin{equation*}
   \widehat f_k^{(\ell)}:= \big(F^{(\ell)}\big)^{-1}\left(\frac k{K+1}\right), \qquad k=1,2, \ldots, K, \quad \ell=1,2,\ldots, L. 
\end{equation*}
Finally, we also apply the stratified sampling approach of \citet{hsa16} and refer to the corresponding ECC variant as {\bf ECC-S}. In this approach
\begin{equation*}
   \widehat f_k^{(\ell)}:= \big(F^{(\ell)}\big)^{-1}(u_k), \qquad k=1,2, \ldots, K, \quad \ell=1,2,\ldots, L,
 \end{equation*}
where \ $u_k$ \ is a random draw from a uniform distribution on \ $\big ]\frac{k-1}K,\frac{k}K\big], \ k=1,2, \ldots ,K$.

Note that in Section \ref{subs4.2} we also investigate the forecast skill of three naive multivariate forecasts obtained from the univariate post-processing methods without accounting for temporal dependencies. These ``independent'' forecasts derived using random sampling from the univariate predictive distributions, considering equidistant quantiles and stratified sampling are denoted by {\bf EMOS-R}, {\bf EMOS-Q} and {\bf EMOS-S}, respectively.

\subsubsection{Dual ensemble copula coupling (dECC)}
\label{subs3.2.2}

The dual ensemble copula coupling (dECC) method of \citet{bbhtp16} combines the structure of the raw ensemble forecast with the estimated forecast error autocorrelation.
\begin{enumerate}
\item Apply ECC-Q to generate an initial post-processed multivariate ensemble forecast \ $\widetilde{\boldsymbol{\mathfrak f}}_1,\widetilde{\boldsymbol {\mathfrak f}}_2, \ldots ,\widetilde{\boldsymbol {\mathfrak f}}_K$ \ with \ $\widetilde{\boldsymbol {\mathfrak f}}_k=\big( \widetilde f_k^{(1)}, \widetilde f_k^{(2)}, \ldots ,  \widetilde f_k^{(L)}\big)^{\top}, \ k=1,2, \ldots K$.
\item With the help of the estimated \ $L\times L$  \ autocorrelation matrix \ $\widehat{\boldsymbol\Sigma}_{\boldsymbol e}$ \ of the forecast error of the ensemble mean generate a correction term
  \begin{equation*}
    \boldsymbol {\mathfrak c}_k:=\widehat{\boldsymbol\Sigma}_{\boldsymbol e}^{1/2}\big(\widetilde{\boldsymbol {\mathfrak f}}_k - \boldsymbol {\mathfrak f}_k\big), \qquad \qquad k=1,2, \ldots ,K,
  \end{equation*}
where \ $\boldsymbol {\mathfrak f}_k=\big(f_k^{(1)}, f_k^{(2)}, \ldots , f_k^{(L)}\big)^{\top}$ \ denotes the $k$th raw multivariate forecast. The estimates of the error correlations can be obtained e.g. from the training data for the univariate post-processing at the different forecast horizons.
\item Derive the adjusted multivariate ensemble  \ $\breve{\boldsymbol{\mathfrak f}}_1,\breve{\boldsymbol {\mathfrak f}}_2, \ldots ,\breve{\boldsymbol {\mathfrak f}}_K$, \ where \ $\breve{\boldsymbol{\mathfrak f}}_k:= \boldsymbol {\mathfrak f}_k + \boldsymbol {\mathfrak c}_k, \ k=1,2, \ldots ,K$.
\item Apply ECC-Q again; however, this time using the rank order structure of the adjusted ensemble of step 3 for rearranging the samples generated from the calibrated univariate predictive distributions.   
\end{enumerate}

\subsubsection{Schaake shuffle (SSh)}
\label{subs3.2.3}

By contrast to ECC, the Schaake shuffle \citep[SSh;][]{cghrw04} determines a dependence template based on past observations rather than the raw ensemble predictions. Samples drawn from the calibrated univariate predictive distributions are thus rearranged in the rank order structure of randomly selected historical observation trajectories of length \ $L$. \  Again, as historical data one can consider the training data for the univariate post-processing. The SSh approach allows for generating post-processed forecasts of any size, provided one has a long enough historical climatological record. However, to ensure a fair comparison with the (d)ECC methods, we restrict the sample size to the size of the raw ensemble. Similar to the ECC described in Section \ref{subs3.2.1}, three different sampling methods from the predictive distributions are considered; the corresponding SSh variants are referred to as {\bf SSh-R} (random sample), {\bf SSh-Q} (equidistant quantiles) and {\bf SSh-S} (stratified sample). In addition to the standard SSh, we further apply two more recently proposed variants that utilize more advanced procedures to select past observations for determining the dependence template.

\subsubsection{Minimum divergence Schaake shuffle (mdSSh)}
\label{subs3.2.4}

The minimum divergence Schaake shuffle \citep[mdSSh;][]{shwhh17} provides a more sophisticated method for selecting historical observation trajectories used as a dependence template. The basic selection algorithm applied for T2M and V10 ensemble forecasts is the following.
\begin{enumerate}
\item For each lead time \ $\ell$, \ calculate the $99\,\%$ central prediction interval of the corresponding marginal predictive distribution  \ $F^{(\ell)}, \ \ell=1,2, \ldots ,L$.
\item From the historical climatological record keep those observation trajectories of length \ $L$, \ where the corresponding central prediction intervals of step 1 contain at least \ $m$ \ observations. Threshold \ $m$ \ is chosen to retain at least \ $M\geq K$ \ forecast trajectories. 
\item Select randomly \ $K$ \ observation trajectories from the \ $M$ \ remaining after step 2.
\end{enumerate}
For the discrete-continuous marginal predictive distributions of PPT24 (see Section \ref{subs3.1.3}), \citet{shwhh17} suggests to replace step 3 by a more complex method to reduce the number of selected forecast trajectories from \ $M$ \ to the required \ $K$. \ This approach is based on selection of a $K$-subset of the set of forecast trajectories of step 2, which minimizes the total divergence of the EMOS predictive CDFs and the empirical CDFs of the corresponding observations \citep[see e.g.][]{tgg13} for all lead times and forecast cases. For more details and a computationally feasible algorithm we refer to \citet{shwhh17}.

\subsubsection{Similarity-based Schaake shuffle (simSSh)}
\label{subs3.2.5}

In contrast to the mdSSh, where observation trajectories used as dependence templates are selected on the basis of their consistency with the corresponding EMOS predictive distributions, the similarity-based Schaake shuffle \citep[simSSh;][]{schefzik16} looks for historical forecast trajectories where the corresponding ensemble forecasts are the most similar to the actual ones.
\begin{enumerate}
\item For a given initialization time in the verification period, calculate the mean \ $\overline{f}^{(\ell)}$ \ and variance \ ${S^2}^{(\ell)}$ \ of the ensemble forecast \ $\boldsymbol f^{(\ell)}$ \ initialized at this time point for each forecast horizon \ $\ell=1,2,\ldots ,L$.
\item For each initialization time \ $\tau$ \ in the historical data set compute similarity
  \begin{equation*}
    \Delta ^{\tau}:=\sqrt{ \sum_{\ell=1}^L\Big(\overline{f}^{(\ell)} - \overline{f}^{(\ell)}_{\tau} \Big)^2 +\frac 1L \sum_{\ell=1}^L\Big({S^2}^{(\ell)} - {S^2_{\tau}}^{(\ell)} \Big)^2 },
  \end{equation*}
  where \ $\overline{f}^{(\ell)}_{\tau}$ \ and \  ${S^2_{\tau}}^{(\ell)}$ \ denote the mean and variance of the ensemble forecast  \ $\boldsymbol f^{(\ell)}_{\tau}$  \ with lead time \ $\ell$ \ initialized at time point \ $\tau$, \ respectively.
\item Chose initialization times \ $\tau_1, \tau_2, \ldots ,\tau_K$ \ resulting in the highest similarity to the actual forecasts, that is where \ $\Delta ^{\tau_1}, \Delta ^{\tau_2}, \ldots ,\Delta ^{\tau_K}$ \ are the smallest among all similarities computed in step 2. The dependence template is given by historic observations \ $y_{\tau_k}^{(\ell)}$ \ corresponding to ensemble forecasts \ \ $\boldsymbol f^{(\ell)}_{\tau_k},  \ k=1,2, \ldots , K, \ \ell=1,2, \ldots ,L$. 
  \end{enumerate}

\subsection{Forecast evaluation methods}
\label{subs3.3}

As argued in \citet{gbr07}, the main goal in probabilistic forecasting is to maximize the sharpness of the predictive distribution subject to calibration. Calibration measures the statistical consistency between the predictions and the corresponding observations, whereas sharpness refers to the concentration of the predictive distribution. Predictive performance is usually quantified with the help of proper scoring rules, which are loss functions \ $\mathcal S(F,y)$ \ assigning numerical values to forecast-observation pairs \ $(F,y)$. \ One of the most popular proper scoring rules in the atmospheric sciences assessing simultaneously both calibration and sharpness is the continuous ranked probability score \citep[CRPS;][Section 9.5.1]{w19}.  For a predictive CDF \ $F(x)$ \  and an observation  \ $y\in\mathbb R$, \ the CRPS is defined as
\begin{equation}
  \label{eq:CRPS}
\crps\big(F,y\big):=\int_{-\infty}^{\infty}\big (F(x)-{\mathbbm 
  I}_{\{x \geq y\}}\big )^2{\mathrm d}x={\mathsf E}|X-y|-\frac 12
{\mathsf E}|X-X'|,
\end{equation}
where \ ${\mathbbm I}_H$ \ denotes the indicator of a set \ $H$, \ whereas \ $X$ \ and \ $X'$ \ are independent random variables with CDF \ $F$ \ and finite first moment. The CRPS is a negatively oriented score, that is the smaller the better, and this scoring rule serves as loss function in parameter estimation of normal, truncated normal and censored generalized extreme value EMOS models for calibration of T2M, V10 and PPT24 ensemble forecasts, respectively.  For these distributions the CRPS can be obtained in closed form \citep[for the corresponding formulae see e.g.][]{jkl19}, allowing a computationally efficient estimation process.

A multivariate extension of the CRPS is the energy score \citep[$\es$;][]{gr07}. Given an $L$-dimensional CDF \ $F$ \ and  vector \ $\boldsymbol y= \big(y^{(1)},y^{(2)}, \ldots ,y^{(L)}\big)^{\top}$, \ the $\es$ is defined as
\begin{equation}
 \label{eq:ES}
\es(F,\boldsymbol y):={\mathsf E}\Vert \boldsymbol X-\boldsymbol
y\Vert-\frac 12 {\mathsf E}\Vert \boldsymbol X-\boldsymbol X'\Vert,
 \end{equation}
 where \ $\Vert \cdot \Vert$ \ denotes the Euclidean distance and, similar to the univariate case \eqref{eq:CRPS}, \ $ \boldsymbol X$ \ and \  $\boldsymbol X'$ \ are independent random vectors having distribution \ $F$. \ For a  forecast ensemble \ $\boldsymbol {\mathfrak f}_1,\boldsymbol {\mathfrak f}_2, \ldots ,\boldsymbol {\mathfrak f}_K$ \ one should consider the empirical CDF \ $F_K$ \ \citep{gsghj08}, which reduces \eqref{eq:ES} to the ensemble energy score
\begin{equation}
  \label{eq:ensES}
\es(F_K,\boldsymbol y)=\frac 1K \sum _{j=1}^K\Vert \boldsymbol
{\mathfrak f}_j-\boldsymbol y\Vert-\frac 1{2K^2}\sum_{j=1}^K\sum_{k=1}^K\Vert
\boldsymbol {\mathfrak f}_j-\boldsymbol {\mathfrak f}_k\Vert.
\end{equation}
Note that the same definition applies for reordered calibrated samples discussed in Section \ref{subs3.2}.

A more recently introduced  multivariate proper scoring rule is the (ensemble) variogram score of order $p$ \citep[$\vs^p$;][]{sh15}. For an ensemble forecast \  $\boldsymbol {\mathfrak f}_k=\big(f_k^{(1)},f_k^{(2)}, \ldots ,f_k^{(L)}\big)^{\top}\!,$ \ $k=1,2, \ldots ,K,$  \ it is defined as
\begin{equation}
  \label{eq:ensVS}
  \vs^p(F_K,\boldsymbol y)=\sum_{i=1}^L\sum_{j=1}^L \omega_{ij}\left(\big| y^{(i)}-y^{(j)}\big|^p  - \frac 1K\sum_{k=1}^K \big| f_k^{(i)}-f_k^{(j)}\big|^p\right)^2,
\end{equation}
where \ $\omega_{ij}\geq 0$ \ is the weight for coordinate pair \ $(i,j)$. \ Compared to the $\es$, the $\vs^p$ is more sensitive to the errors in the specification of correlations. Following \citet{lbm20}, we consider here \ $p=1$ \ and use the notation \ $\vs$ \ for \ $\vs^1$.

Further, in the case studies of Section \ref{sec4}, for a given forecast \ $F$ \ the improvement in terms of a score  \ $\mathcal S_F$ \ with respect to a reference forecast  \ $F_\text{ref}$ \ is quantified using the corresponding skill score \citep{gr07}
\begin{equation*}
    \label{eq:skillScore}
  {\mathcal S}^{skill}_F :=1-\frac{\overline{\mathcal S}_F}{\overline{\mathcal S}_{F_\text{ref}}}, 
\end{equation*}
where \ $\overline{\mathcal S}_F$ \ and \ $\overline{\mathcal S}_{F_\text{ref}}$ \ denote the mean score values over all forecast cases in the verification period for forecasts \ $F$ \ and  \ $F_\text{ref}$, \ respectively. Thus, besides the $\es$ and the $\vs$ we investigate the energy skill score ($\ess$) and the variogram skill score (VSS), which are positively oriented (the larger the better).

Calibration of univariate ensemble forecasts can also be diagnosed with the help of verification rank histograms displaying the ranks of observations with respect to the ensemble forecasts \citep[see e.g.][Section 9.7.1]{w19}. For a properly calibrated $K$-member ensemble the ranks follow a uniform distribution on \ $\{1,2, \ldots ,K+1\}$, \ and the deviation from uniformity can be quantified by the reliability index 
\begin{equation}
   \label{eq:relind}
 \ri:=\sum_{r=1}^{K+1}\Big| \rho_r-\frac 1{K+1}\Big|,
\end{equation}
where \ $\rho_r$ \ is the relative frequency of rank \ $r$ \ \citep{dmhzds06}. There are several options to generalize the verification rank histogram to multivariate ensemble forecasts depending on the definition of the ranks in higher dimensions. Here we consider the average ranking given by the average of the univariate ranks of the different coordinates. The resulting histogram has similar properties and interpretation as the univariate rank histogram \citep{tsh16}.

Further, we also investigate the mean Euclidean distance (EE) of the median vectors of the forecasts from the corresponding validating observations, where the multivariate $L^1$ ensemble median can be obtained e.g. with the algorithm of \citet{vz00}. 

Finally, following \citet{gr11}, statistical significance of score differences between forecasts are assessed with the help of the Diebold-Mariano \citep[DM;][]{dm95} test, which is able to account for temporal dependencies in the forecast errors.  Given a scoring rule \ $\mathcal S$ \ and two competing probabilistic forecasts \ $F$ \  and \ $G$, \ the test statistic of the DM test is given by
\begin{equation}
  \label{eq:DM}
  t_N = \sqrt{N} \frac{\overline{\mathcal S}_F - \overline{\mathcal S}_G}{\widehat \sigma_N}, 
\end{equation}
where \ $\overline{\mathcal S}_F$ \ and \ $\overline{\mathcal S}_G$ \ are the mean scores over a test set corresponding to forecasts \ $F$ \ and \ $G$, \ respectively, and \ $\widehat \sigma_N$ \ is a suitable estimator of the asymptotic standard deviation of the sequence of individual score differences.  Under standard regularity conditions, \ $t_N\ $ asymptotically follows a standard Gaussian distribution under the null hypothesis of equal predictive performance. Negative values of \ $t_N$ \ indicate a better predictive performance of \ $F$, \  whereas \ $G$ \  is preferred in the case of positive values of \ $t_N$.

\section{Results}
\label{sec4}

The predictive performance of the different multivariate post-processing methods is investigated in three case studies based on global ECMWF T2M, V10 and PPT24 ensemble forecasts. To assess calibration of probabilistic forecasts we use the energy score ($\es$), the variogram score of order 1 ($\vs$) and we also investigate the multivariate rank histograms together with the corresponding reliability indices ($\Delta$), whereas multivariate point forecasts are evaluated with the help of the mean Euclidean error (EE).

While our focus is on the multivariate predictive performance, the first step in each of the multivariate methods described in Section \ref{subs3.2} is the independent calibration of ensemble forecasts with different forecast horizons. Hence, we start with a brief description of the details of univariate post-processing, followed by the results for the multivariate models.

\subsection{Univariate post-processing}
 \label{subs4.1}
For a given initialization time, the corresponding ensemble forecasts with different forecast horizons are calibrated with the help of the EMOS approaches of Section \ref{subs3.1}. Estimates of the EMOS model parameters minimize the mean CRPS of the predictive distribution over locally selected training data using a rolling training period. This means that for each observation station, the training data set consists of ensemble forecasts and corresponding validating observations of the given station for the preceding \ $n$ \ calendar days. To avoid numerical problems in the minimization of the loss function, local models require long training periods; here we consider the optimal training period lengths determined by \citet{hspbh14}.

 \subsubsection{Temperature}
 \label{subs4.1.1}

\begin{figure}[t]
   \centering
   \hbox {\epsfig{file=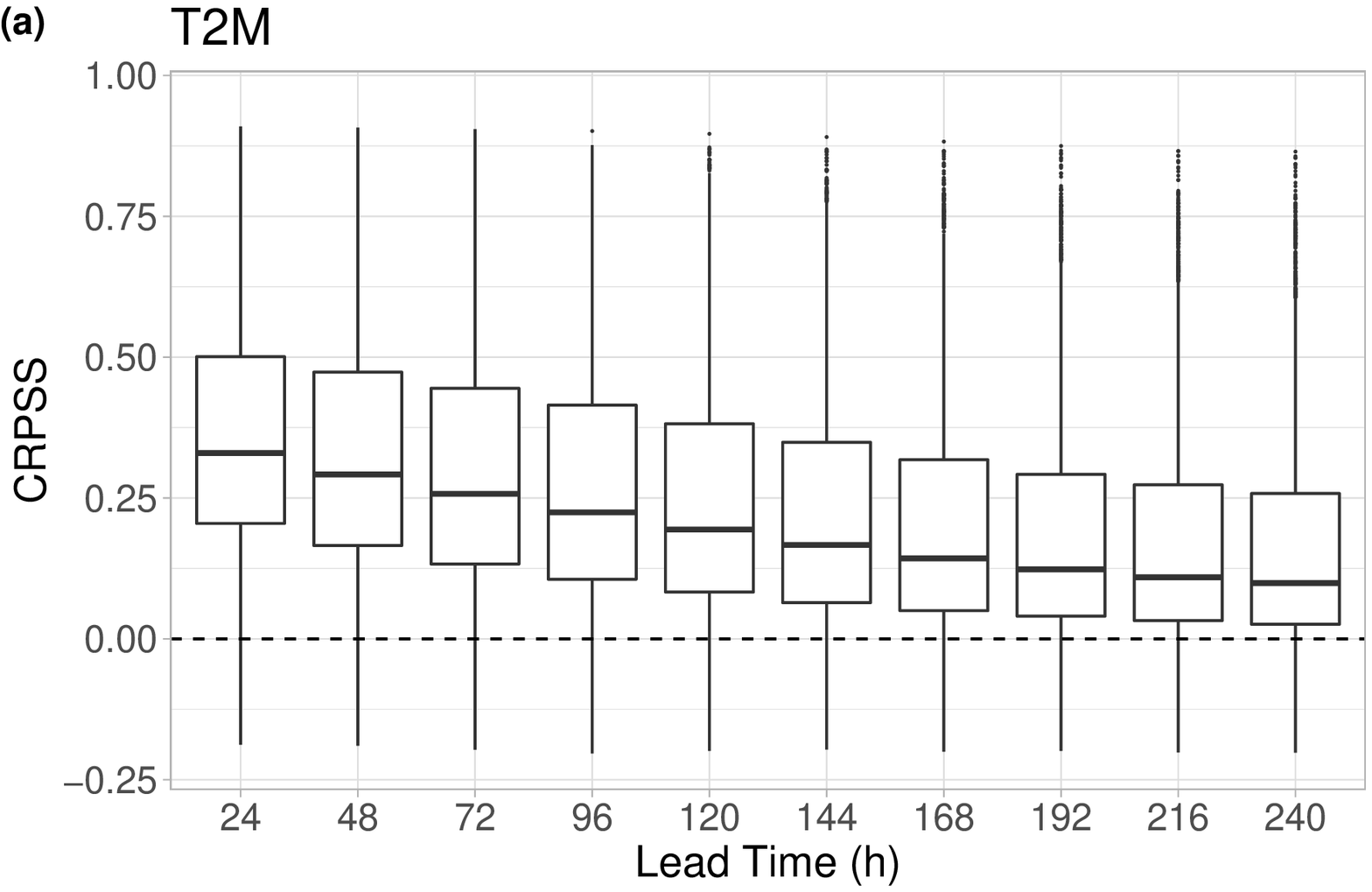, width=.5\textwidth} \epsfig{file=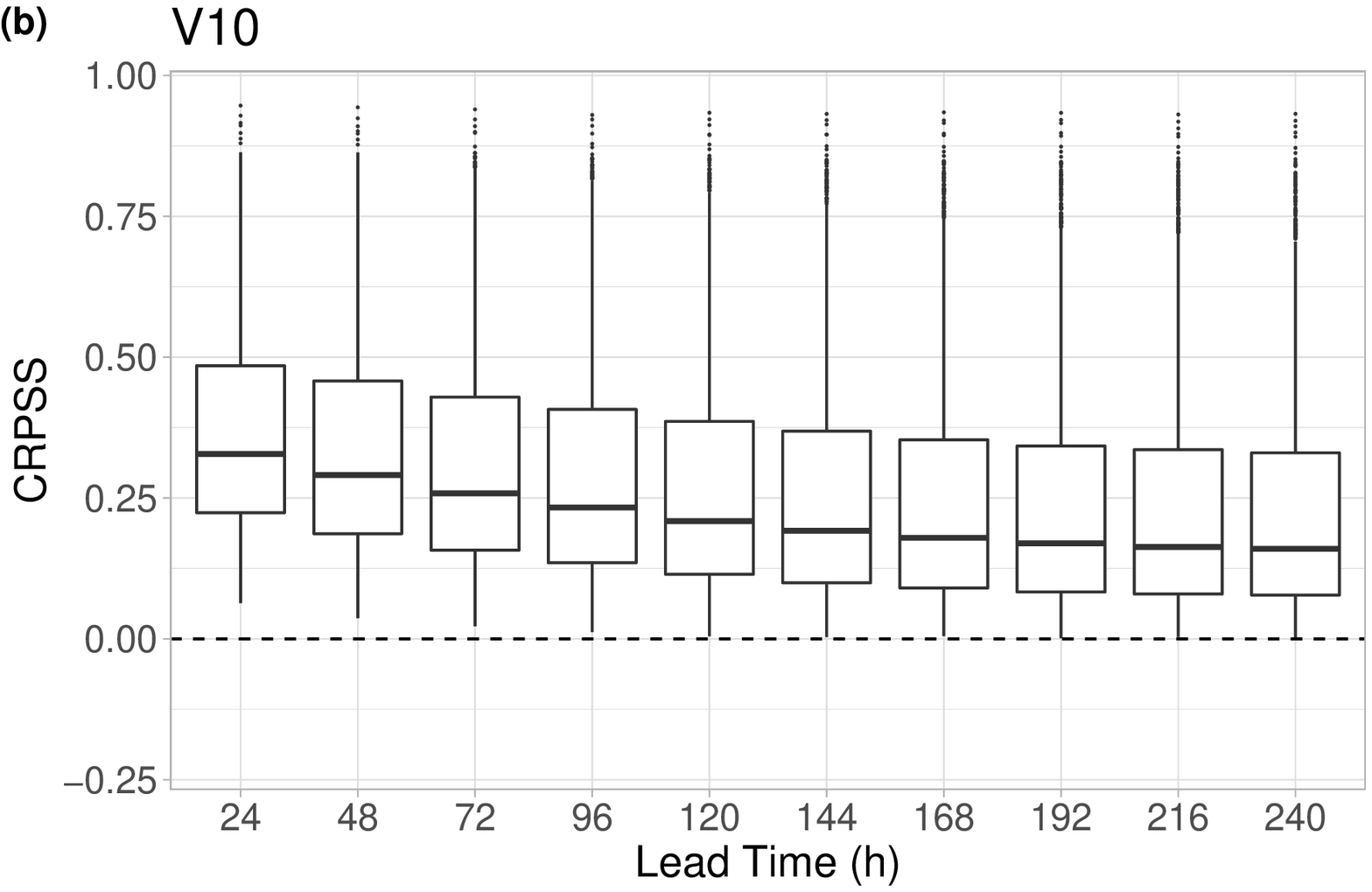, width=.5\textwidth}}

   \medskip
   \epsfig{file=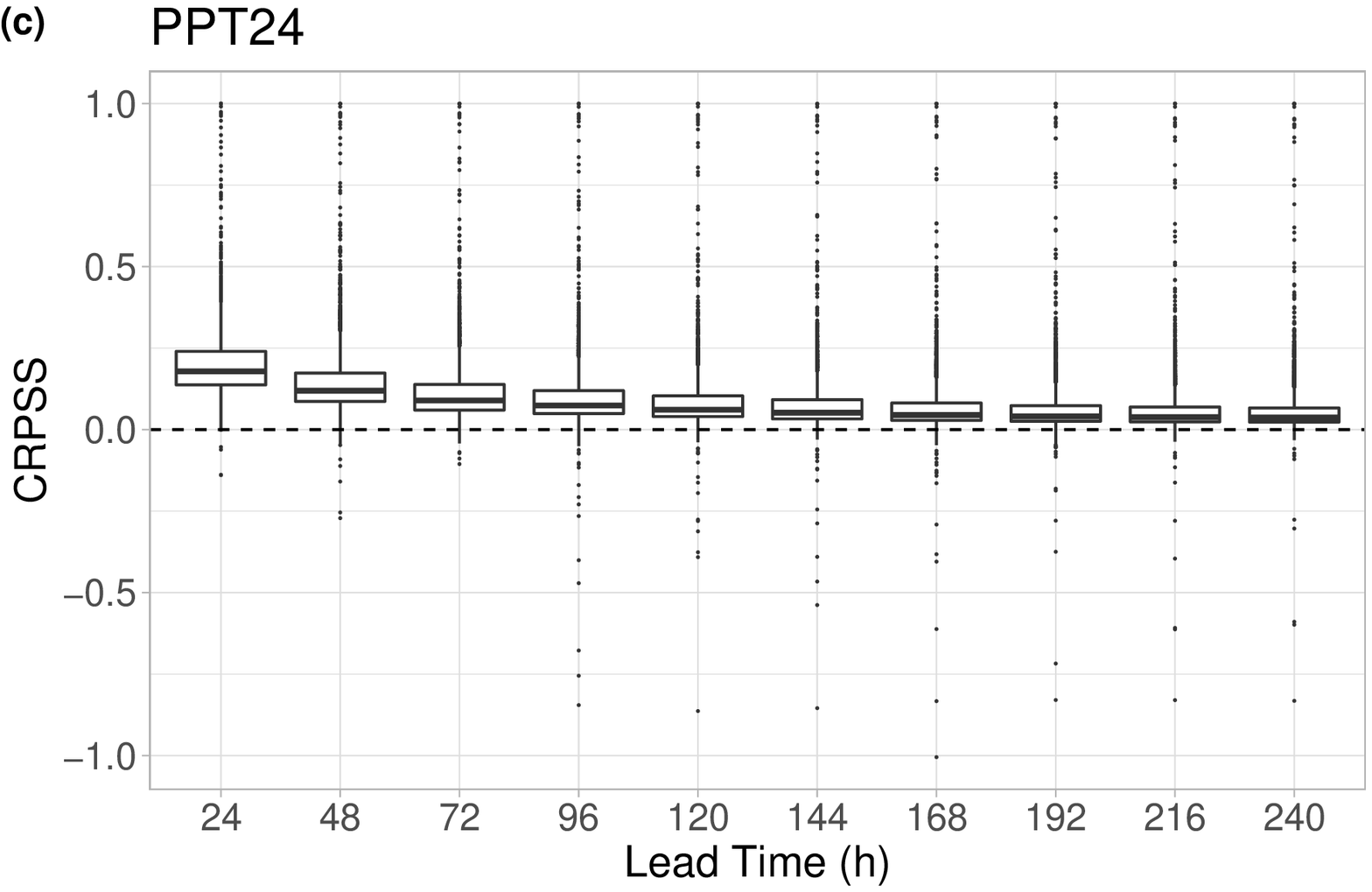, width=.5\textwidth}

   \caption{Boxplots of the CRPSS of the EMOS models for T2M (a), V10 (b) and PPT24 (c) over the verification period with respect to the corresponding raw ensemble forecasts.}
   \label{fig:crpss}
 \end{figure}

 $\ell$-day ahead ECMWF T2M ensemble forecasts \ ($\ell =1,2,\ldots 10$) \ are calibrated with the help of the normal EMOS model \eqref{emosT2M} with a training period length of $720$ days. For verification we consider the time interval between 1 January 2004 and 20 March 2014 (3732 calendar days). Figure \ref{fig:crpss}a shows the boxplots of the CRPSS of the EMOS models over the verification period with respect to the raw ensemble forecasts.
Compared with the raw ECMWF T2M forecasts, univariate post-processing substantially decreases the mean CRPS for the vast majority of stations for all lead times; however, the gain decreases with the increase of the forecast horizon \citep[see also][]{frg19}.

  \subsubsection{Wind speed}
  \label{subs4.1.2}
The optimal training period length for the truncated normal EMOS model \eqref{emosV10} for post-processing ECMWF V10 ensemble forecasts is 365 calendar days. This allows an almost one-year longer verification period than in the case of T2M; however, to keep the results consistent, we verify both models on the same time interval starting at 1 January 2004. The forecast skill of the EMOS models for various lead times in terms of the mean CRPS over the verification period is depicted in Figure \ref{fig:crpss}b. Compared to the T2M ensemble forecasts, EMOS post-processing results in a positive skill score for all considered stations and the improvement in the mean CRPS is a bit higher, especially for longer lead times.

\subsubsection{Precipitation}
\label{subs4.1.3}

Precipitation records usually contain a large number of zero observations, hence, reliable parameter estimation in EMOS modelling requires much longer training periods than for temperature or wind speed. Following \citet{hspbh14}, to calibrate ECMWF PPT24 forecasts with lead times of \ $1,2, \ldots ,10$ \ days, we make use of the EMOS model \eqref{emosPPT24} with a 1816-day rolling training period leaving the time interval from 1 January 2007 to 20 March 2014 for model verification. Again, in Figure \ref{fig:crpss}c we provide the CRPSS values for various forecast horizons, where, similar to the other two weather quantities, the reference forecast is the raw ensemble. The general behaviour of the skill scores is the same as before. For almost all stations EMOS forecasts outperform the raw PPT24 ensemble for all lead times and the improvement decreases with the increase of the forecast horizon. However, compared with T2M and V10, boxplots of the CRPSS values of the EMOS models for PPT24 have much shorter interquartile ranges (IQRs) and display far more outliers.

  \subsection{Multivariate performance}
  \label{subs4.2}
We now continue with the comparison of the forecast skill of the multivariate approaches described in Section \ref{subs3.2} using calibrated samples of size $52$, which is the size of the raw ECMWF ensemble (see Section \ref{sec2}). In methods requiring historical data for providing the dependence template (estimation of the autocorrelation matrix for dECC; observation  trajectories for the SSh and for its variants), we consider forecast-observation pairs from the rolling training window applied in univariate calibration. For T2M and V10 ensemble forecasts, in mdSSh we select dependence template from  observation trajectories of the training data set where at least \ $m=6$ \ from the \ $L=10$ \ observations falls into the corresponding central prediction interval, whereas in the case of PPT24, due to the high computational costs \citep[see also][]{shwhh17}, this method is excluded from the analysis.

\begin{figure}[th!]
   \centering

   \epsfig{file=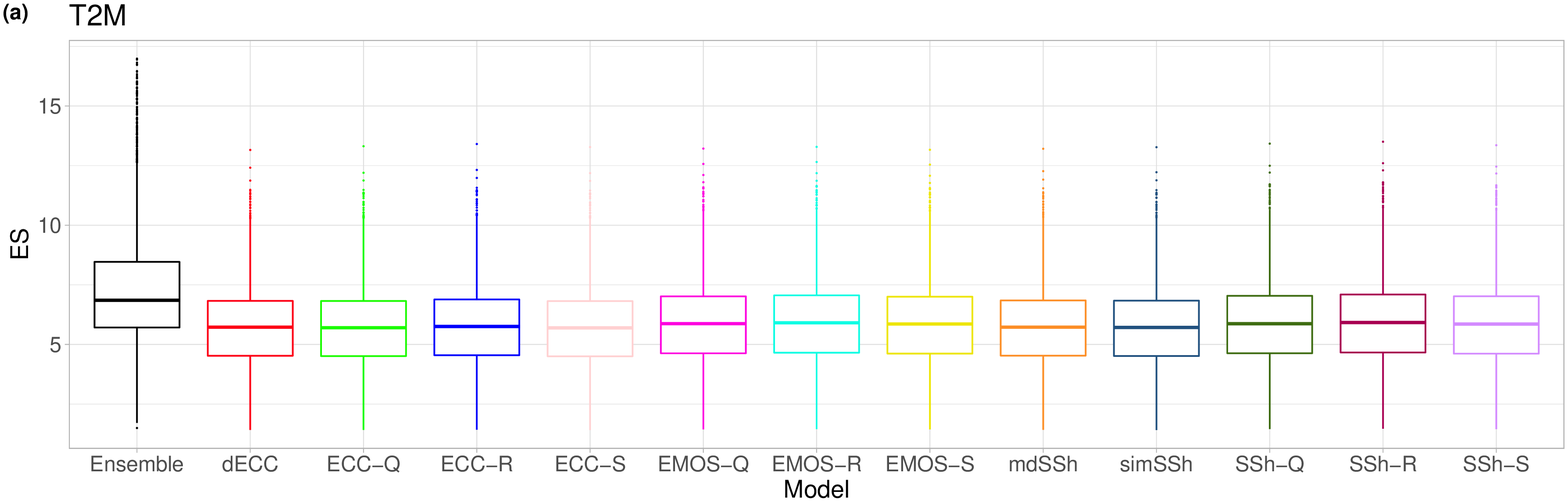, width=\textwidth}

   \smallskip
   \epsfig{file=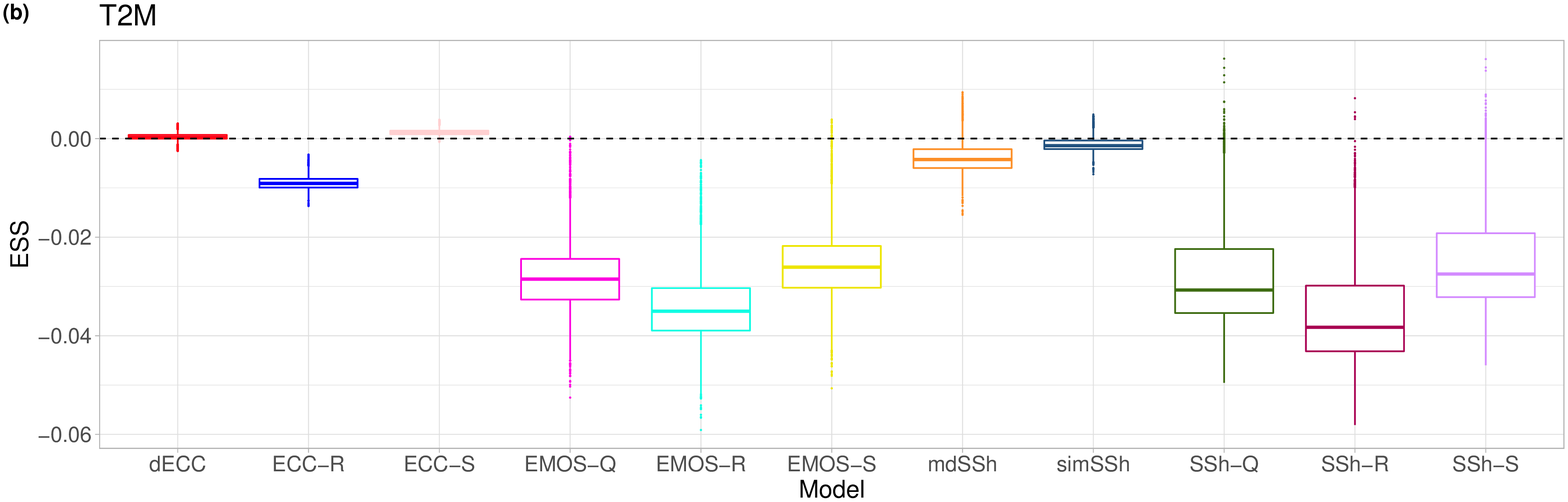, width=\textwidth}

   \smallskip
    \epsfig{file=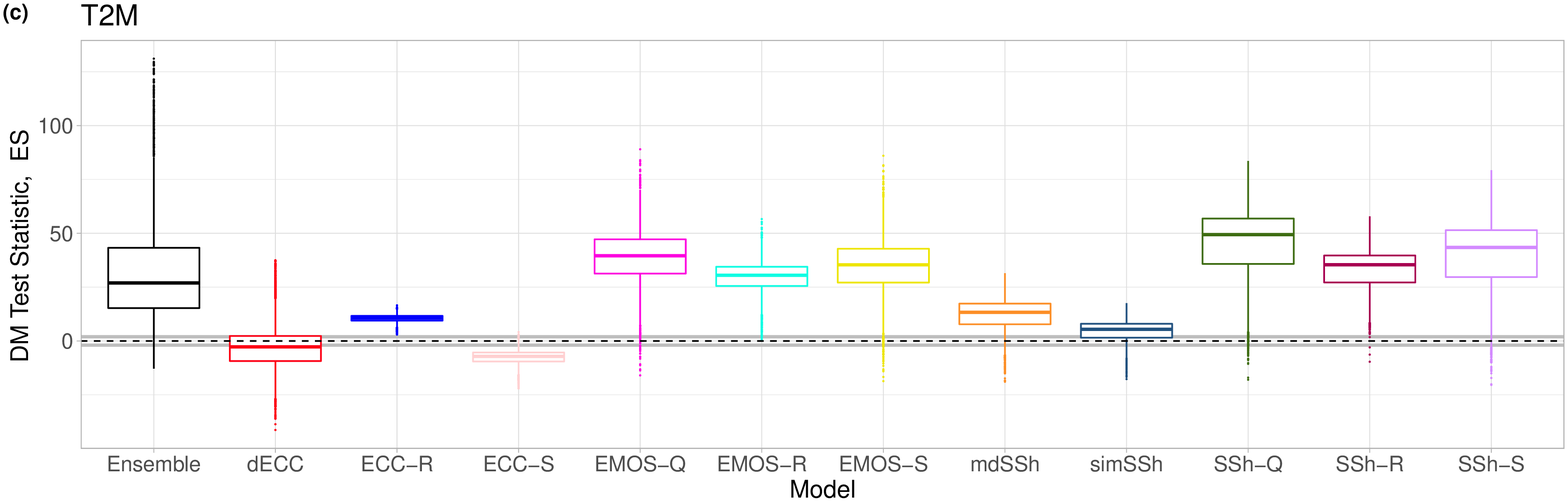, width=\textwidth}

    \caption{Boxplots of the mean ES over the verification period of the calibrated and raw T2M forecasts (a); of the ESS with respect to the ECC-Q approach (b); of the DM test statistic investigating the significance of the difference from the reference ECC-Q method (c). Grey lines indicate the acceptance region of the two-tailed DM test for equal predictive performance at a 5\,\% level of significance. 
}
   \label{fig:t2m_ES}
 \end{figure}

In the following analysis, the ECC-Q forecast is used as a reference for the computation of skill scores and in DM tests for equal predictive performance (see Section \ref{subs3.3}). Further, in the interest of improving the visual presentation of the results, in the boxplots presented in Sections \ref{subs4.2.1} -- \ref{subs4.2.3} extreme values (further than 3$\cdot$IQR from the box) are not indicated.

  \subsubsection{Temperature}
  \label{subs4.2.1} 

 \begin{figure}[th!]
   \centering

   \epsfig{file=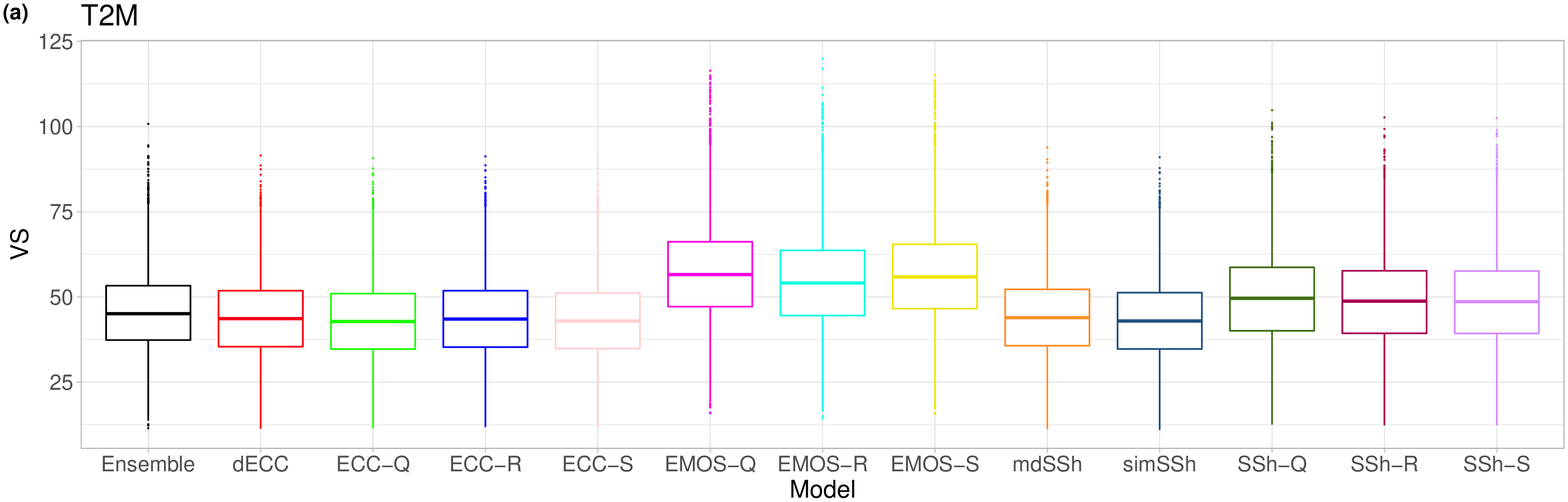, width=\textwidth}

   \smallskip
   \epsfig{file=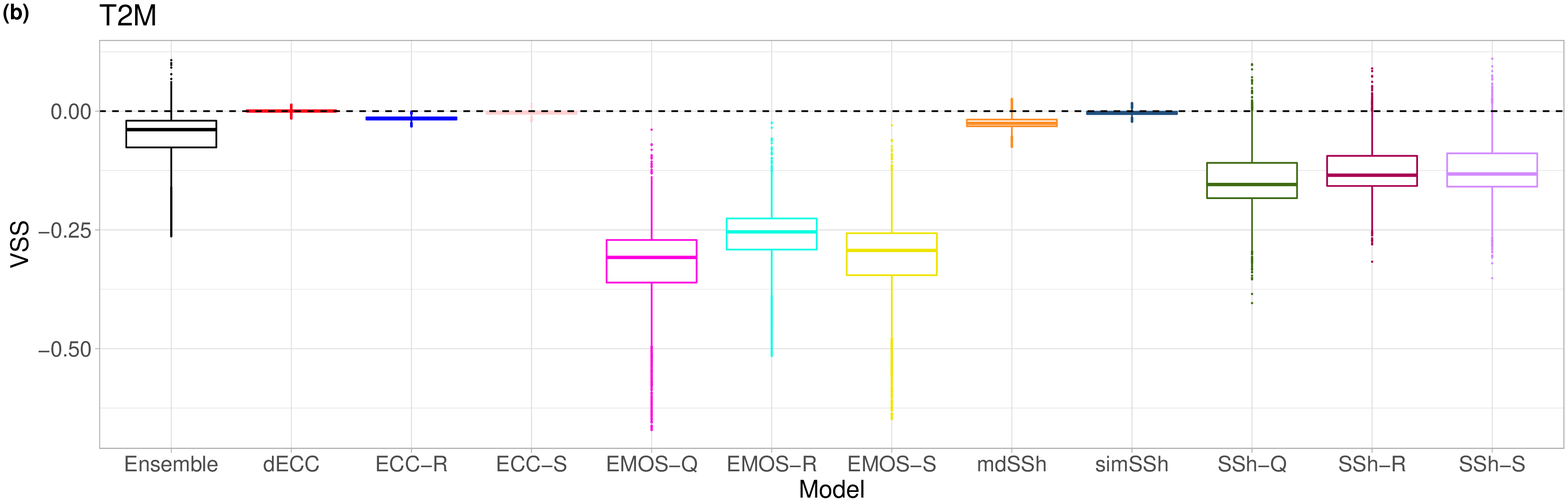, width=\textwidth}

     \smallskip
    \epsfig{file=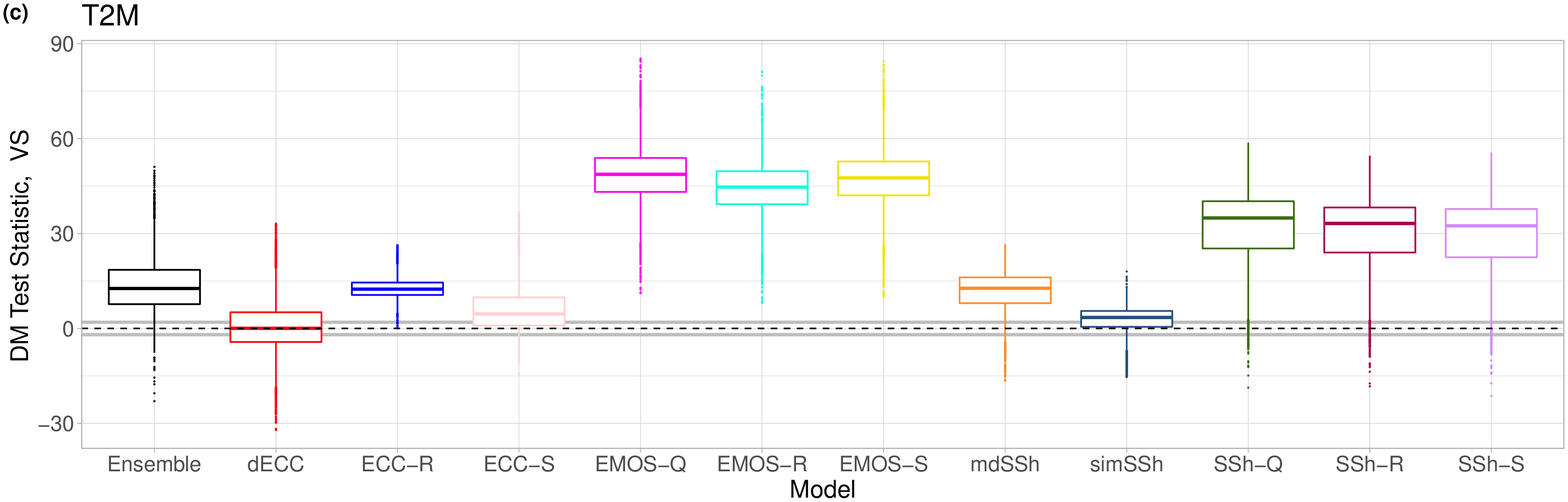, width=\textwidth}

   \caption{As Figure \ref{fig:t2m_ES}, but summarizing the results for the VS.}
   \label{fig:t2m_VS}
 \end{figure}

 In Figure \ref{fig:t2m_ES}, the various multivariate post-processing approaches and the raw T2M forecasts are compared in terms of the mean ES \eqref{eq:ensES}  of the different observation stations over the verification period (1 January 2004 -- 20 March 2014). According to Figure \ref{fig:t2m_ES}a, each calibration method improves the $\es$ of the raw ensemble; however, the differences between the various approaches are hardly visible. A better insight can be obtained from Figure \ref{fig:t2m_ES}b displaying the $\ess$ values with respect to the reference ECC-Q forecast. The raw ensemble forecasts resulting in very low skill scores are excluded here. ECC-S shows the best predictive performance, closely followed by ECC-Q and dECC. Perhaps surprisingly, even the advanced variants of the Schaake shuffle (mdSSh and simSSh) fail to outperform the reference ECC-Q method, whereas the standard SSh with any of the three investigated sampling schemes is behind in skill and fail to provide clear improvements over the EMOS models that do not account for multivariate dependencies. Finally, according to the results of the DM tests for equal predictive performance (Figure \ref{fig:t2m_ES}c), the differences in $\es$ from the reference model are significant for all approaches but the dECC.

\begin{figure}[th!]
   \centering

   \epsfig{file=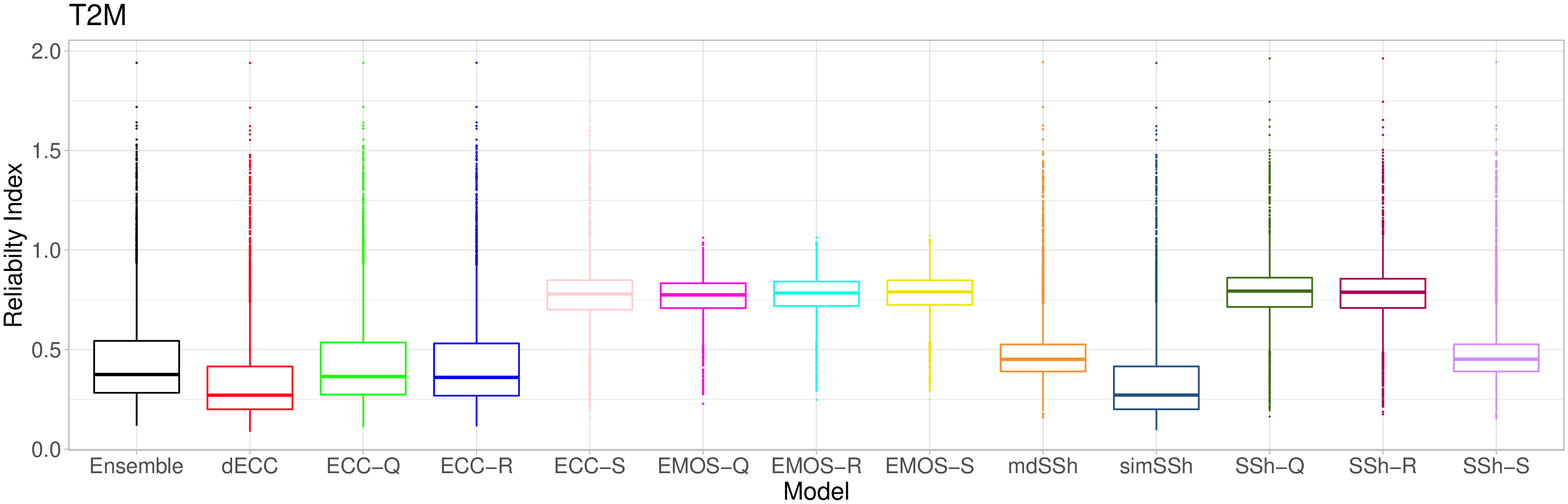, width=\textwidth}

    \caption{Boxplots of the reliability indices corresponding to average ranks over the verification period of the calibrated and raw T2M forecasts.}
   \label{fig:t2m_RI}
 \end{figure}
 
 A similar ranking of the multivariate post-processing methods can be observed in Figure \ref{fig:t2m_VS}. ECC-Q results in the lowest mean $\vs$; however, its advantage over dECC, ECC-S and simSSh is often not significant (see Figure \ref{fig:t2m_VS}c). Further, in contrast to the $\es$, neither the independent EMOS forecasts, nor the variants of the standard SSh outperform the raw ECMWF ensemble in terms of $\vs$.

\begin{figure}[th!]
   \centering

   \epsfig{file=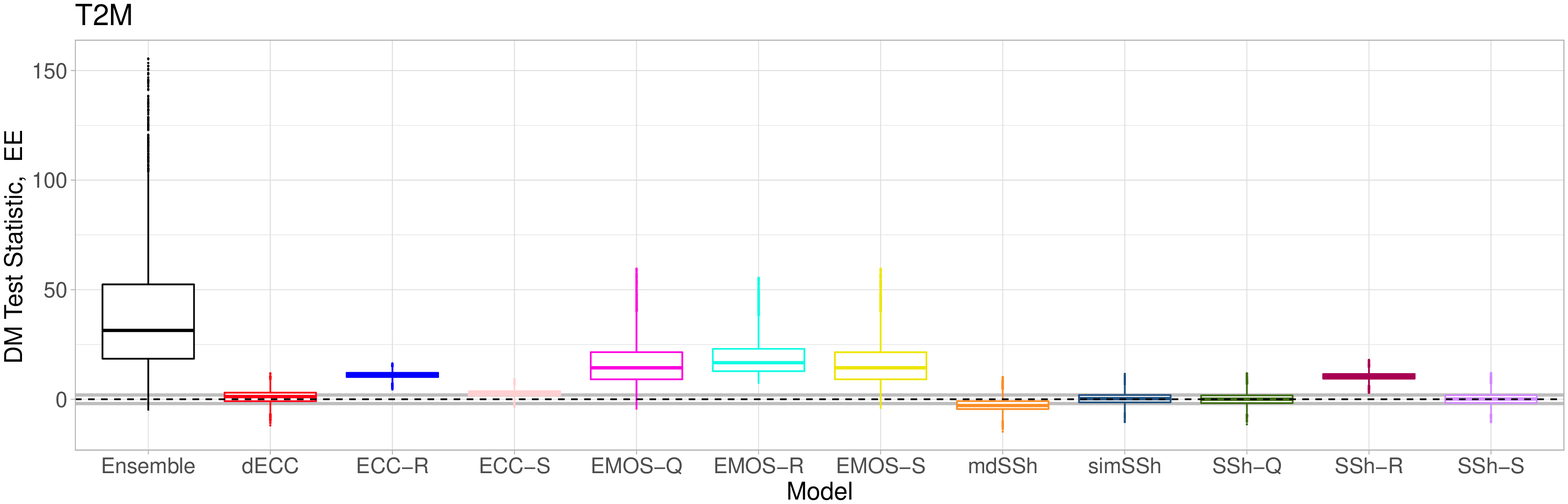, width=\textwidth}

    \caption{Boxplots of the DM test statistic investigating the significance of the difference from the reference ECC-Q method in terms of the mean EE of the $L^1$ median vectors of calibrated and raw T2M forecasts. Grey lines indicate the acceptance region of the two-tailed DM test for equal predictive performance at a 5\,\% level of significance. }
   \label{fig:t2m_L1}
 \end{figure}

A slightly different picture can be obtained from the boxplots of the reliability indices corresponding to average ranks over the verification period of the calibrated and raw T2M forecasts displayed in Figure \ref{fig:t2m_RI}. ECC-S, all three independent predictions (EMOS-Q, EMOS-R, EMOS-S), SSh-Q and SSh-S underperform the raw ensemble by a wide margin, which is a consequence of the highly overdispersive character of these forecasts resulting in hump-shaped rank histograms (nor shown). SSh-S still shows some overdispersion, whereas the other forecasts, including the raw ensemble, are underdispersive. The average ranks of the dECC and simSSh are the closest to the uniform distribution (the corresponding mean/median reliability indices are $0.357/0.271$ for both forecasts), followed by the mdSSh ($0.425/0.347$), ECC-R ($0.438/0.360$) and ECC-Q ($0.442/0.365$); however, the rank histograms of the latter two forecasts (not shown) are just slightly less underdispersive than that of the ECMWF ensemble ($0.451/0.375$).

\begin{figure}[th!]
   \centering

   \epsfig{file=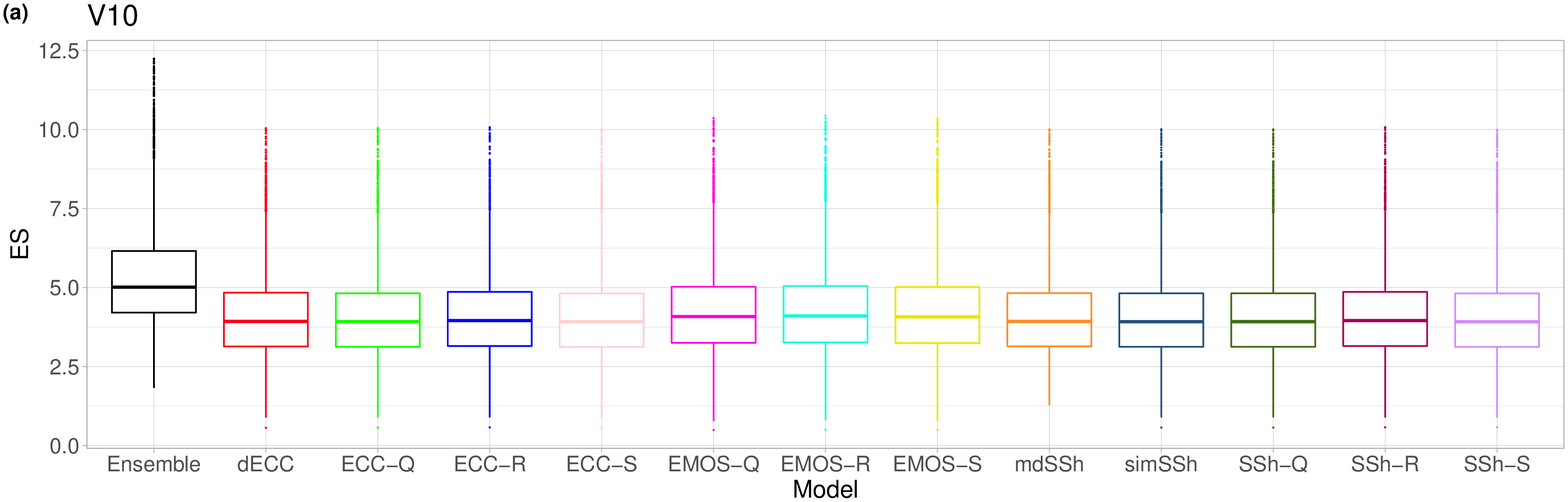, width=\textwidth}

   \smallskip
   \epsfig{file=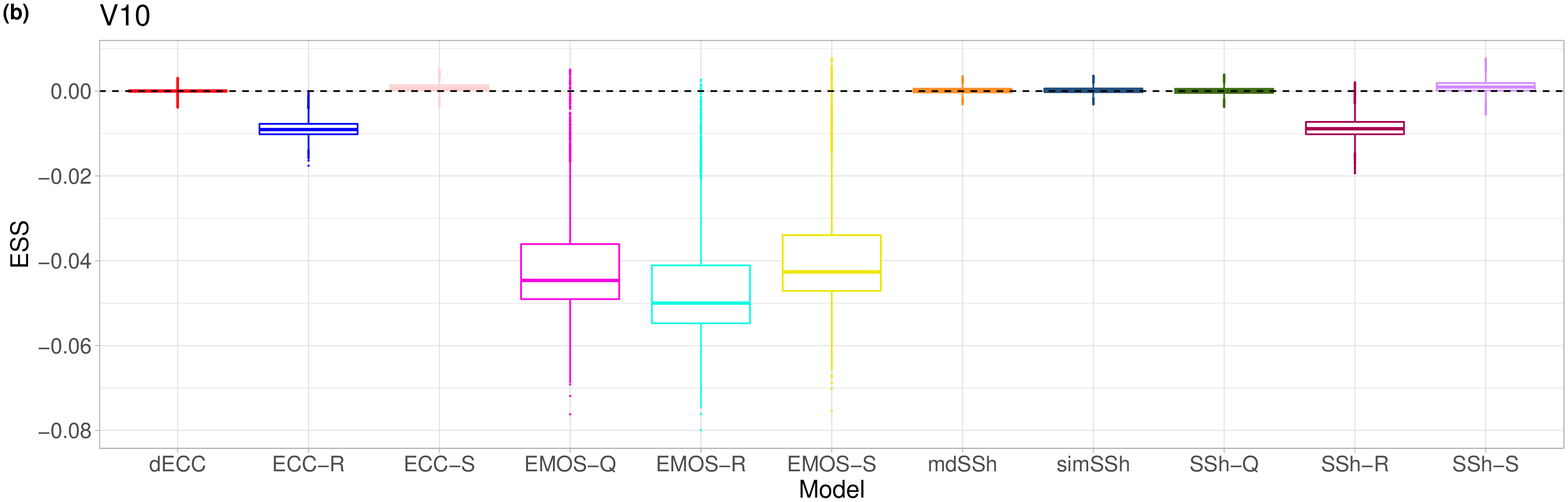, width=\textwidth}

     \smallskip
    \epsfig{file=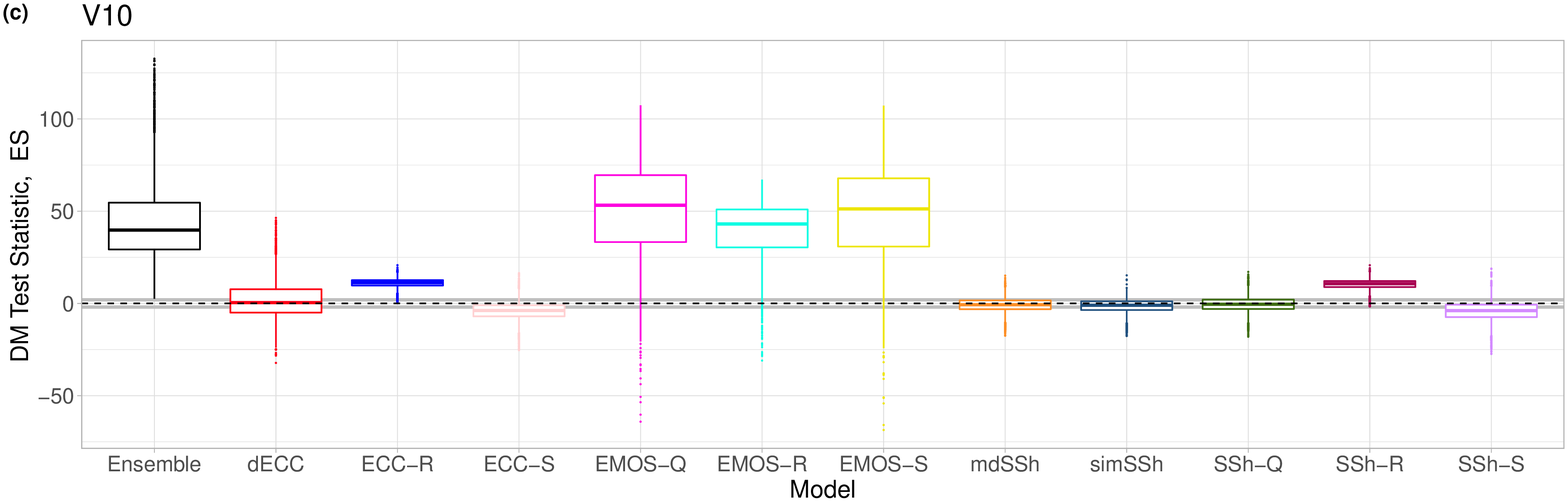, width=\textwidth}

    \caption{Boxplots of the mean ES over the verification period of the calibrated and raw V10 forecasts (a); of the ESS with respect to the ECC-Q approach (b); of the DM test statistic investigating the significance of the difference from the reference ECC-Q method (c). Grey lines indicate the acceptance region of the two-tailed DM test for equal predictive performance at a 5\,\% level of significance.} 
   \label{fig:v10_ES}
 \end{figure}

Finally, Figure \ref{fig:t2m_L1} shows the boxplots of the DM test statistic investigating the significance of the difference from the reference ECC-Q method in terms of the mean EE of the $L^1$ median vectors. Compared with the raw ensemble, post-processing substantially improves the accuracy of the $L^1$ median forecast and the empirical copula-based models clearly outperform the independent approaches. The lowest mean EE corresponds to the mdSSh; however, the differences between this method and the dECC, ECC-Q, ECC-S, simSSh, SSh-Q and SSh-S are not significant.

 \begin{figure}[th!]
   \centering

   \epsfig{file=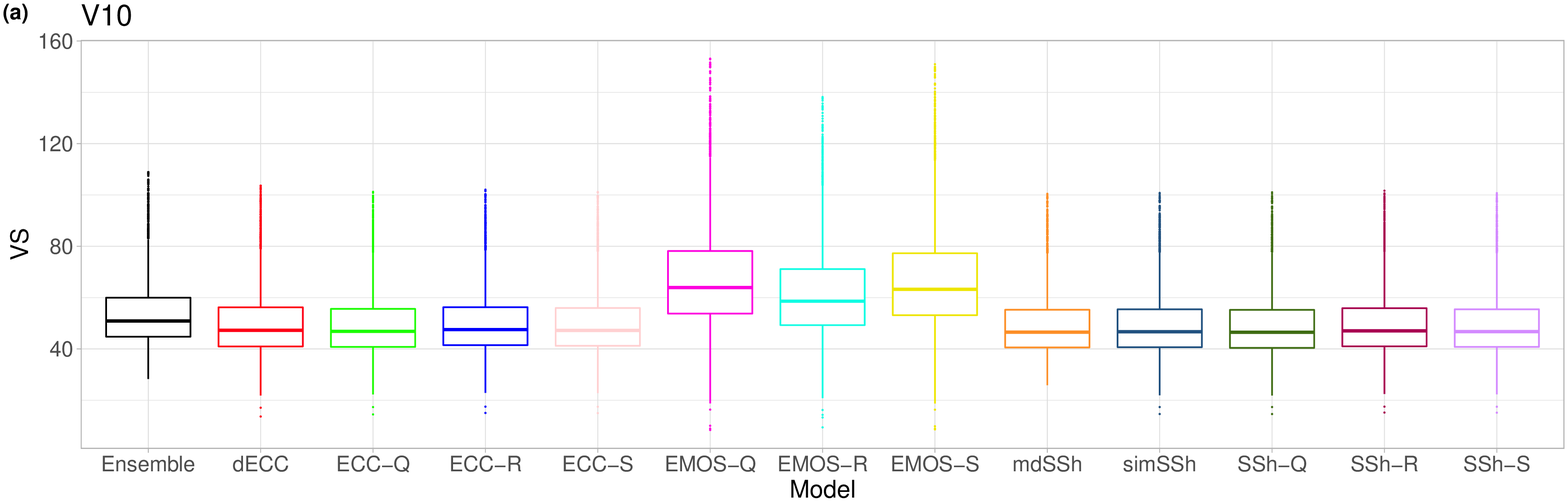, width=\textwidth}

   \smallskip
   \epsfig{file=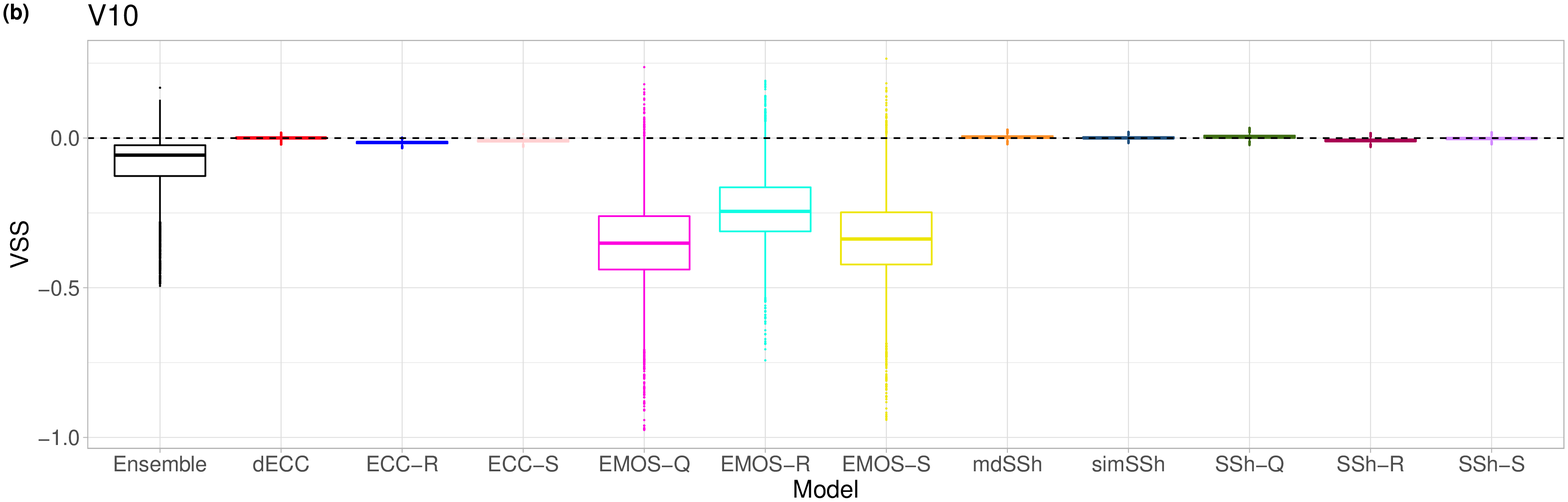, width=\textwidth}

       \smallskip
    \epsfig{file=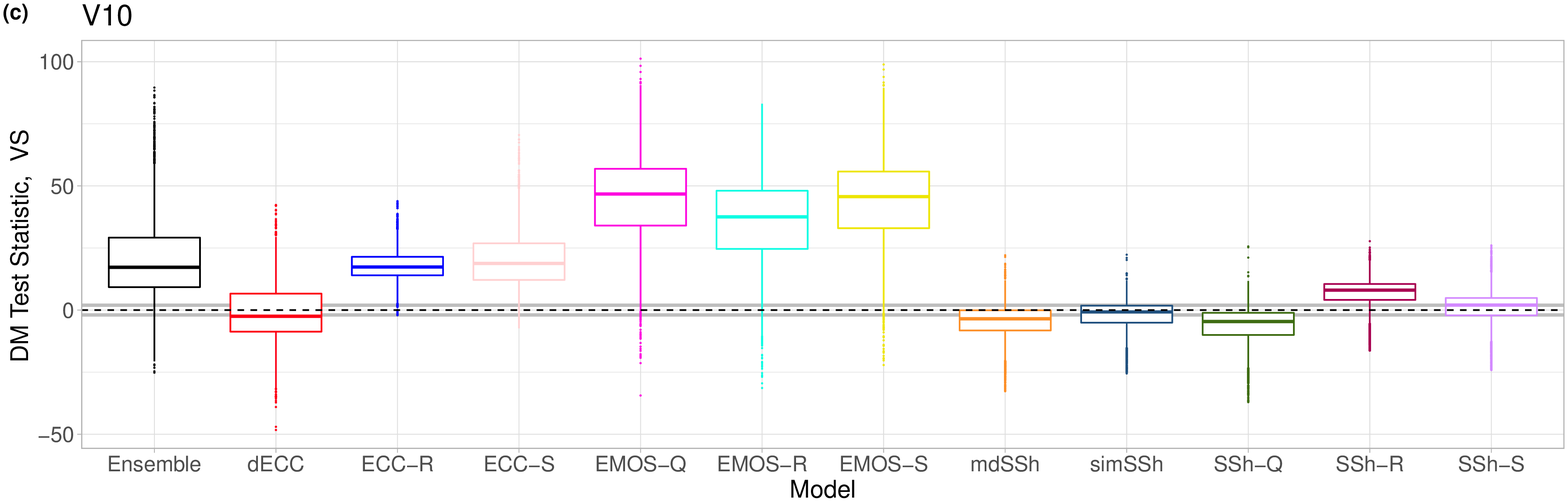, width=\textwidth}

   \caption{As Figure \ref{fig:v10_ES}, but summarizing the results for the VS.}
   \label{fig:v10_VS}
 \end{figure}

\subsubsection{Wind speed}
  \label{subs4.2.2}
 
 The predictive performance of the post-processed and raw V10 ensemble forecast vectors in terms of the mean ensemble $\es$ over the verification period can be investigated with the help of Figure \ref{fig:v10_ES}. Compared with the raw ensemble, each of the investigated post-processing approaches substantially reduces the mean energy score (see Figure \ref{fig:v10_ES}a). According to the skill scores with respect to the ECC-Q model depicted in Figure \ref{fig:v10_ES}b (the raw ensemble is again excluded), the best performing methods are ECC-S and SSh-S, followed by ECC-Q, mdSSh, simSSh and SSh-Q. However, as the DM test statistics provided in Figure \ref{fig:v10_ES}c indicate, the forecast skill of the latter four approaches does not differ significantly. Finally, in contrast to the case of T2M ensemble forecast vectors, all empirical copula-based methods outperform the independent EMOS-Q, EMOS-R and EMOS-S forecasts by a wide margin.

\begin{figure}[t]
   \centering

   \epsfig{file=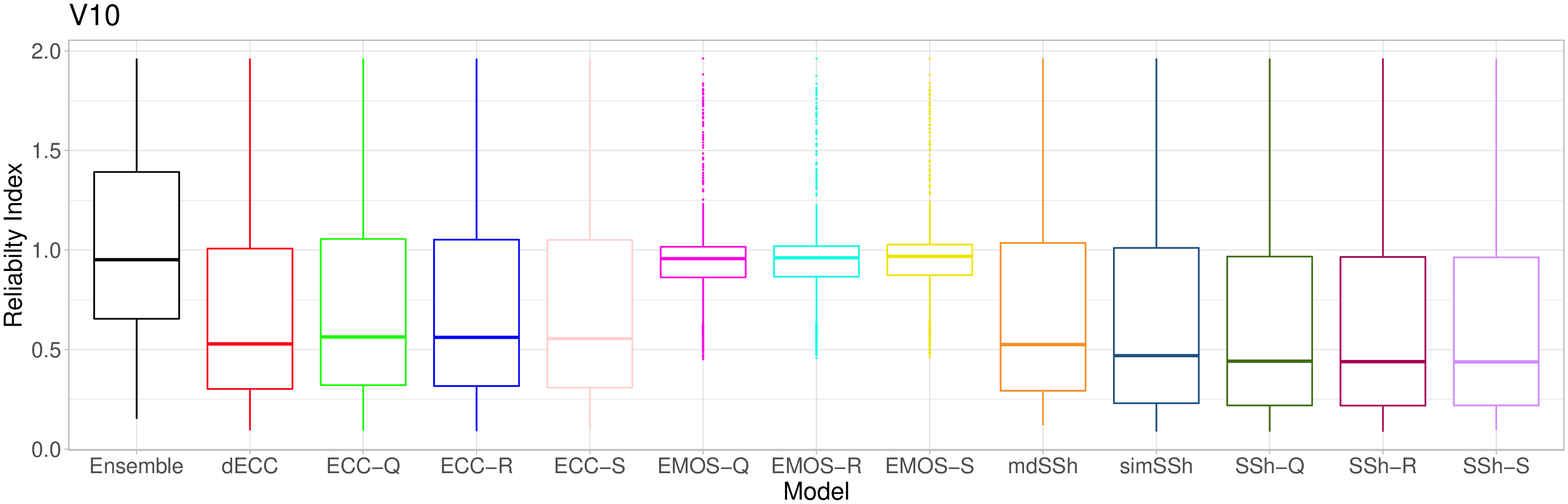, width=\textwidth}

    \caption{Boxplots of the reliability indices corresponding to average ranks over the verification period of the calibrated and raw V10 forecasts.}
   \label{fig:v10_RI}
 \end{figure}

\begin{figure}[t]
   \centering

   \epsfig{file=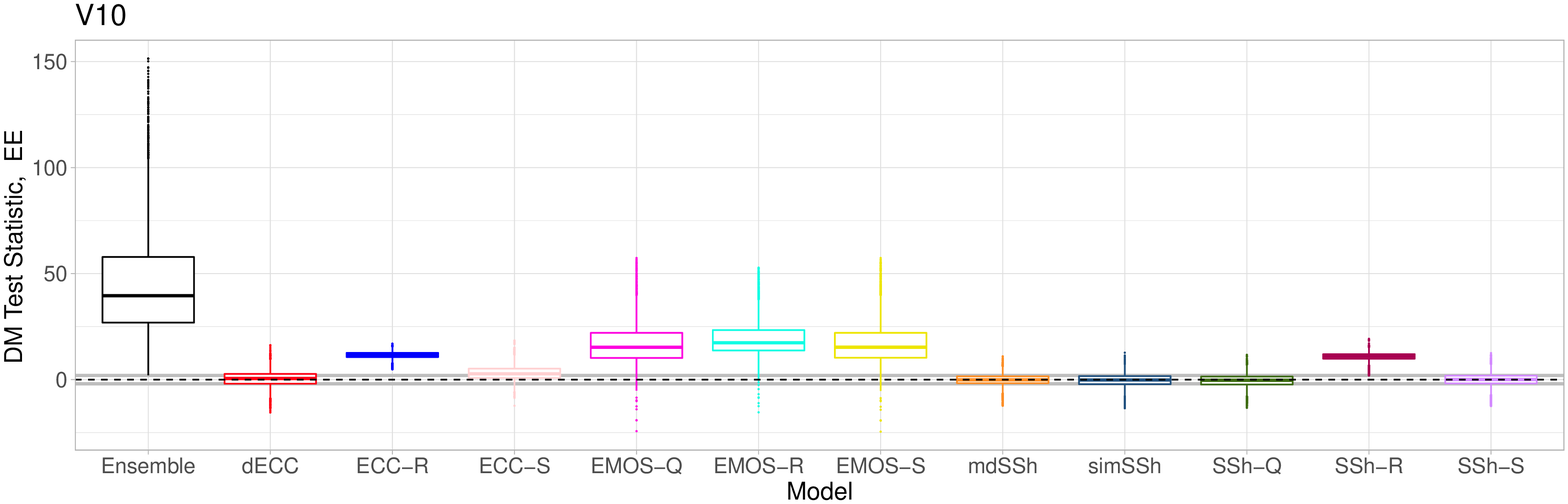, width=\textwidth}

    \caption{Boxplots of the DM test statistic investigating the significance of the difference from the reference ECC-Q method in terms of the mean EE of the $L^1$ median vectors of calibrated and raw V10 forecasts. Grey lines indicate the acceptance region of the two-tailed DM test for equal predictive performance at a 5\,\% level of significance. }
   \label{fig:v10_L1}
 \end{figure} 
 
Figure \ref{fig:v10_VS} summarizing the results for the mean $\vs$ is similar to Figure \ref{fig:t2m_VS} in the sense that independent post-processing of raw V10 forecasts with different lead times (EMOS-Q, EMOS-R, EMOS-S) increases the score values. In general, the predictive performance of the empirical copula-based methods is almost identical (see Figure \ref{fig:v10_VS}b), with only ECC-R, ECC-S and SSh-R performing slightly worse. However, according to the results of the DM tests given in Figure \ref{fig:v10_VS}c, the difference in skill of these three methods from the reference ECC-Q approach is significant.

\begin{figure}[th!]
   \centering

   \epsfig{file=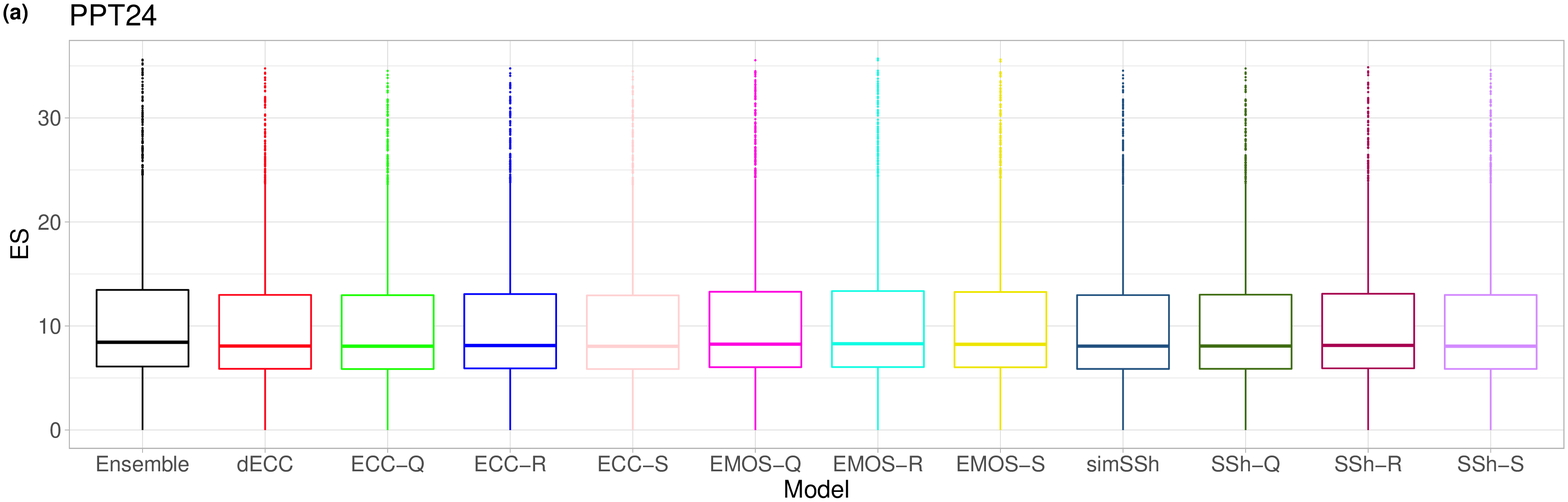, width=\textwidth}

   \smallskip
   \epsfig{file=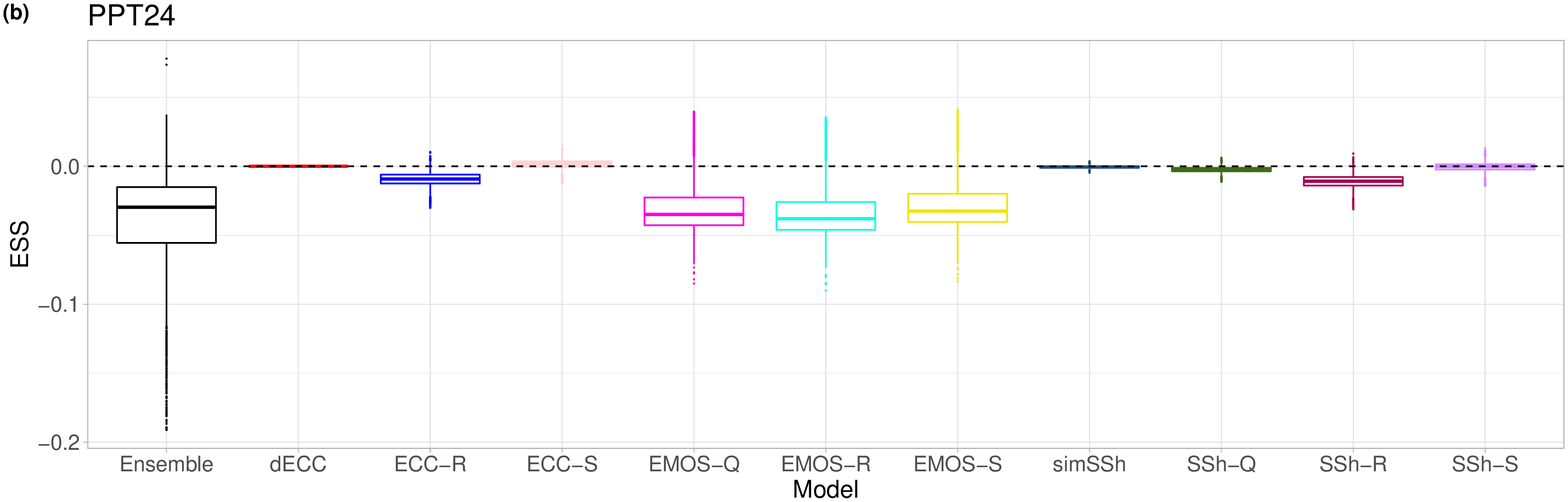, width=\textwidth}

        \smallskip
    \epsfig{file=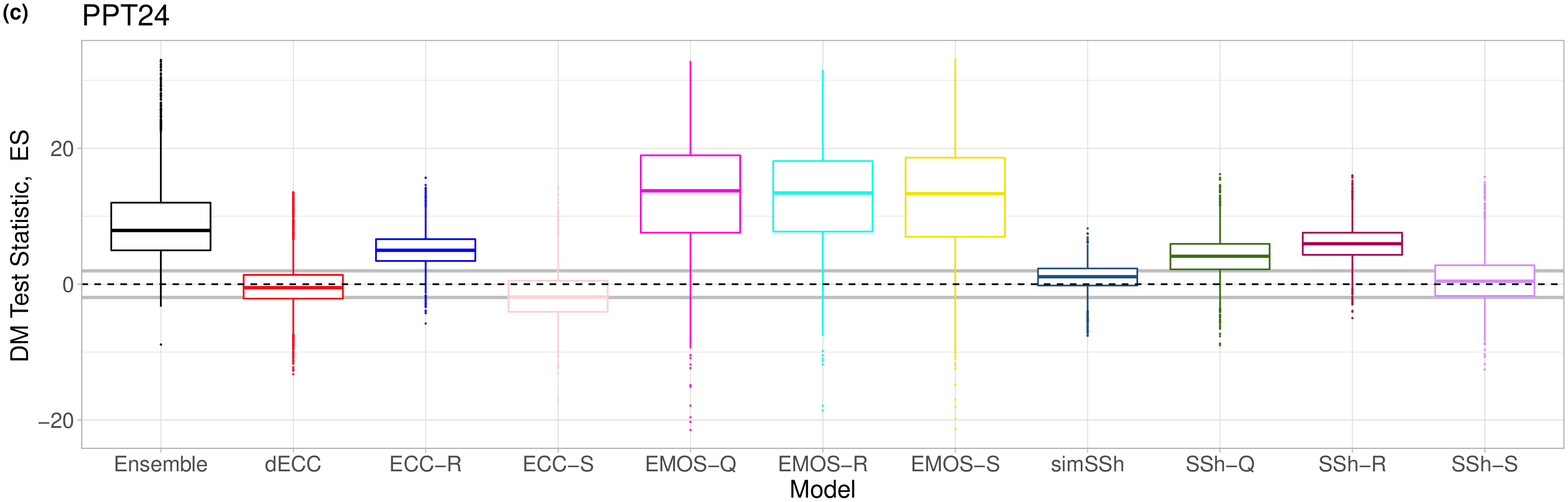, width=\textwidth}

    \caption{Boxplots of the mean ES over the verification period of the calibrated and raw PPT24 forecasts (a); of the ESS with respect to the ECC-Q approach (b); of the DM test statistic investigating the significance of the difference from the reference ECC-Q method (c). Grey lines indicate the acceptance region of the two-tailed DM test for equal predictive performance at a 5\,\% level of significance. 
}
   \label{fig:ppt24_ES}
 \end{figure}

\begin{figure}[th!]
   \centering

   \epsfig{file=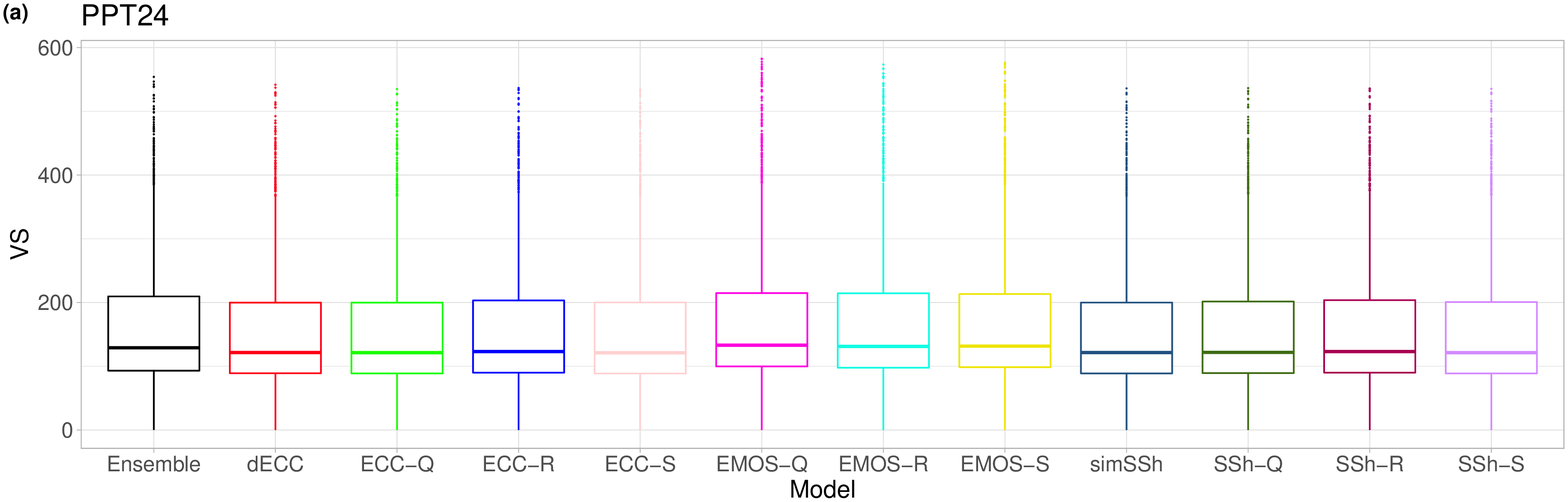, width=\textwidth}

   \smallskip
   \epsfig{file=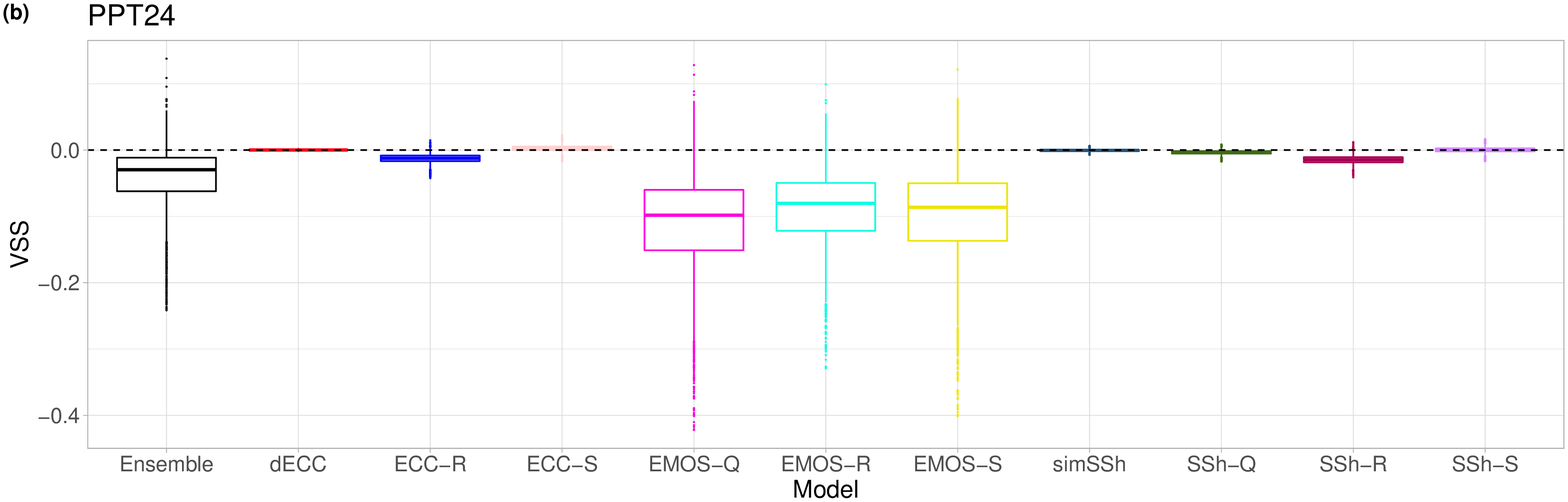, width=\textwidth}

    \smallskip
    \epsfig{file=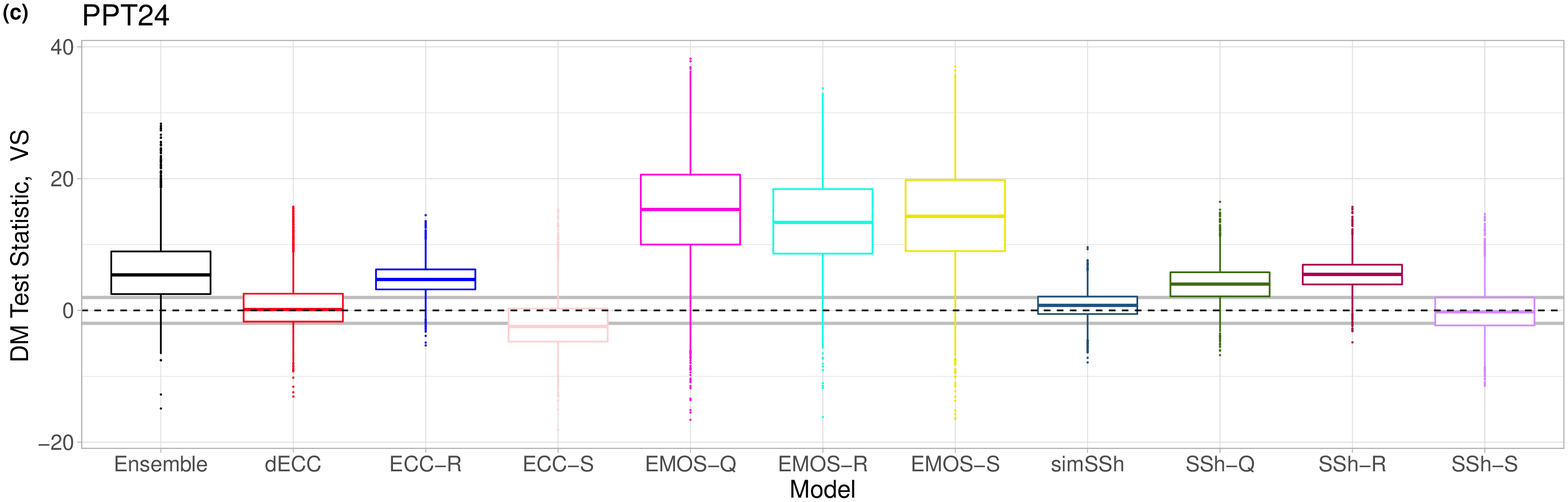, width=\textwidth}

   \caption{As Figure \ref{fig:ppt24_ES}, but summarizing the results for the VS.}
   \label{fig:ppt24_VS}
 \end{figure}

In contrast to temperature, in the case of wind speed any form of post-processing improves the multivariate calibration in the sense that the corresponding average rank histogram is closer to the uniform distribution than the rank histogram of the raw V10 ensemble. This improvement is quantified in the reliability indices displayed in Figure \ref{fig:v10_RI}, where the highest values again correspond to the independently calibrated EMOS-Q, EMOS-R and EMOS-S forecasts. These approaches result in hump-shaped rank histograms (not shown) indicating overdispersion, the corresponding mean/median $\ri$ values are $0.936/0.957$, $0.939/0.961$ and $0.947/0.968$, respectively. All other forecasts (including the raw ensemble) are underdispersive, where the least U-shaped rank histograms correspond to the three versions of the standard Schaake shuffle, followed by the simSSh. The mean/median reliability indices of these forecasts are $0.639/0.441$ (SSh-Q), $0.637/0.439$ (SSh-R), $0.637/0.438$ (SSh-S) and $0.671/0.470$ (simSSh), which are highly under the corresponding scores of $1.029/0.952$ of the raw ensemble vector. 
 
Finally, Figure \ref{fig:v10_L1} showing boxplots of the DM test statistic investigating the significance of the difference from the reference ECC-Q method in terms of the mean EE of the $L^1$ median vectors is almost identical to Figure \ref{fig:t2m_L1}. All post-processing methods result in more accurate median forecasts than the raw ensemble and all empirical copula-based methods but the ECC-R and the SSh-R perform almost identically well.

  \subsubsection{Precipitation accumulation}
  \label{subs4.2.3}

\begin{figure}[th!]
   \centering

   \epsfig{file=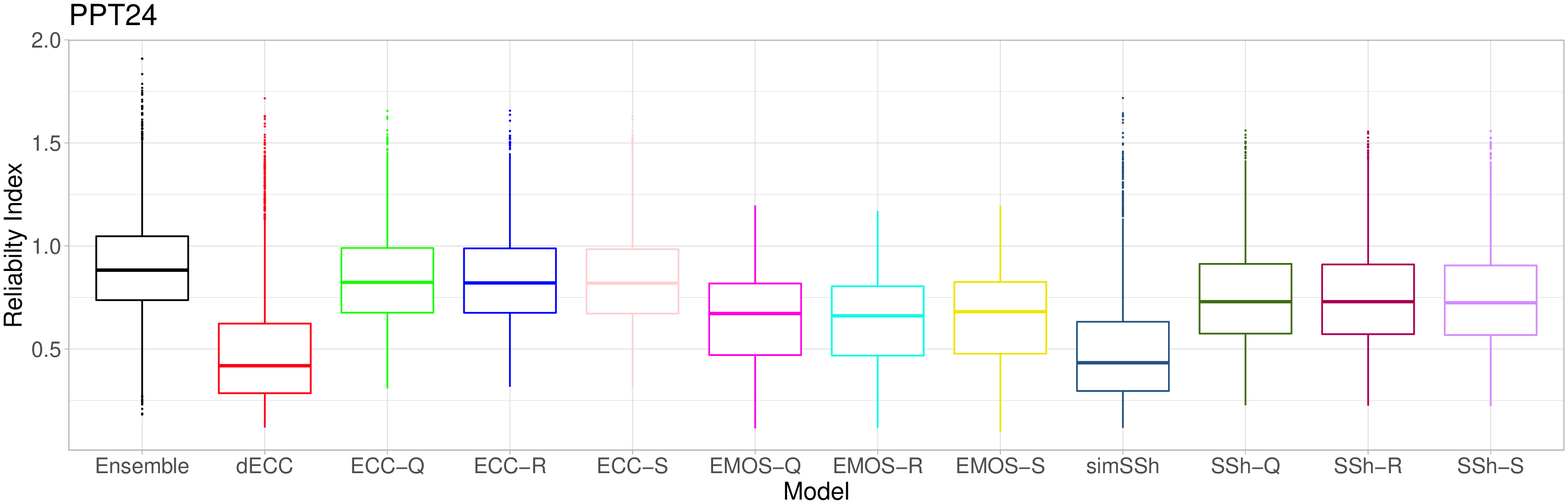, width=\textwidth}

    \caption{Boxplots of the reliability indices corresponding to average ranks over the verification period of the calibrated and raw PPT24 forecasts.}
   \label{fig:ppt24_RI}
 \end{figure}  

Figure \ref{fig:ppt24_ES}a displaying the boxplots of the mean ES over the verification period of the calibrated and raw PPT24 forecasts does not reveal a clearly visible difference between the various predictions. The raw ensemble and the independent EMOS-Q, EMOS-R and EMOS-S methods seem to be slightly behind the other forecasts, which is confirmed by the skill scores of Figure \ref{fig:ppt24_ES}b. From the empirical copula-based approaches ECC-S again shows the best forecast skill, followed by  ECC-Q, dECC, simSSh and SSh-S. However, as indicated by the values of the DM test statistics summarized in  Figure \ref{fig:ppt24_ES}c, the differences between the latter four forecasts in terms of the ES are not  significant.

\begin{figure}[t]
   \centering

   \epsfig{file=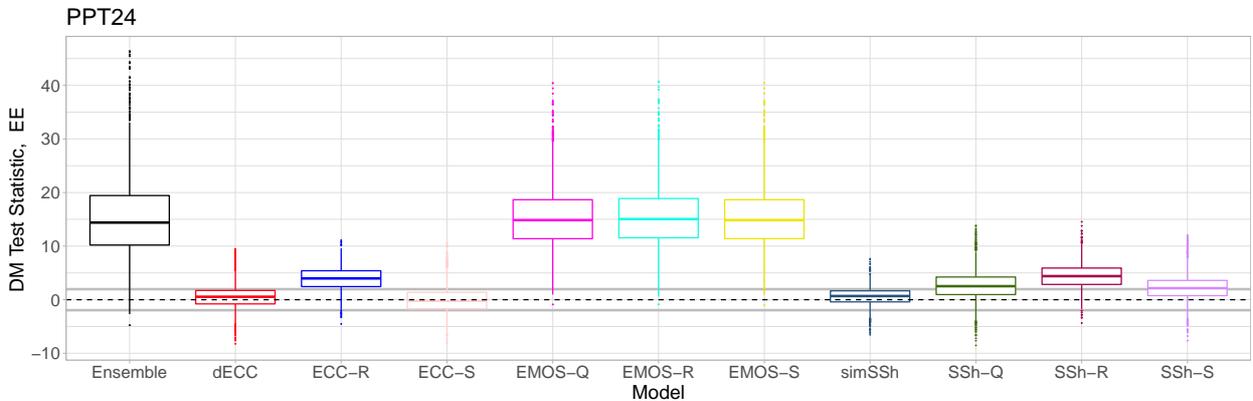, width=\textwidth}

    \caption{Boxplots of the DM test statistic investigating the significance of the difference from the reference ECC-Q method in terms of the mean EE of the $L^1$ median vectors of calibrated and raw PPT24 forecasts. Grey lines indicate the acceptance region of the two-tailed DM test for equal predictive performance at a 5\,\% level of significance. }
   \label{fig:ppt24_L1}
 \end{figure}  

 In contrast to the other two weather quantities,  Figures \ref{fig:ppt24_ES} and  \ref{fig:ppt24_VS} result in the same conclusions, that is the rankings of the forecasts in terms of the ES and the VS are identical.

Further, investigating the boxplots of Figure \ref{fig:ppt24_RI}, one can again observe the positive effect of post-processing on multivariate calibration quantified in lower reliability indices. The average ranks of dECC and simSSh fit most the uniform distribution, followed by the independent EMOS-Q, EMOS-R and EMOS-S approaches. Note that despite their overdispersive character resulting in hump-shaped rank histograms (not shown), these independent forecasts outperform all versions of the standard SSh and ECC, which are highly underdispersive.

Finally, Figure \ref{fig:ppt24_L1} showing the boxplots of the DM test statistic investigating the significance of the difference from the reference ECC-Q method in terms of the mean EE of the $L^1$ median vectors suggests the same ranking of the various post-processing approaches as Figures \ref{fig:ppt24_ES} and \ref{fig:ppt24_VS}. However, in contrast to the ES and VS, with regard to this score even the independent EMOS-Q, EMOS-R and EMOS-S methods outperform the raw ensemble.

\section{Discussion and conclusions}
\label{sec5}

We compared a wide variety of state-of-the-art methods for multivariate ensemble post-processing with a focus on dependencies over lead times from 1 to 10 days, using three case studies of global ECMWF ensemble forecasts of temperature, wind speed and precipitation accumulation. Across all of the three settings all multivariate post-processing methods substantially improve all investigated aspects of multivariate forecast quality over both the raw ensemble predictions as well as a simple application of univariate post-processing models without accounting for multivariate dependencies. Among the basic ECC and SSh variants, the different sampling strategies only showed minor differences in the predictive performance. In particular, random sampling (ECC-R and SSh-R) generally performed worse than quantile-based (ECC-Q and SSh-Q) or stratified sampling (ECC-S and SSh-S), whereas no significant differences could be detected between the latter two approaches. Comparing the more advanced dECC, mdSSh and simSSh approaches with their basic counterparts, we generally did not observe any benefits of these more complex methods, and did not find a single case where they significantly outperform ECC-Q. The only notable exceptions to all of these findings is the case of temperature forecasts where the basic SSh approaches performed surprisingly bad and failed to outperform the raw ensemble predictions or the EMOS model forecasts that do not account for any dependencies. However, this phenomenon can be explained by the stronger seasonal changes in the temporal dependency patterns of temperature, compared to the other two investigated random variables. Random sampling from past observations might thus result in historical observation trajectories of completely different characteristics coming e.g.\ from other seasons. This is nicely corrected with the use of the more advanced selection techniques of mdSSh and simSSh. For instance, in the case study of Section \ref{subs4.2.1}, the median proportion of historical observation trajectories used for providing dependence templates selected from the same season as the prediction to be reshuffled is around 23.0\,\% for the basic SSh methods and 43.4\,\% for the mdSSh, whereas the corresponding upper quartiles are 23.4\,\% and 58.2\,\%, respectively. This finding indicates that in the case of weather quantities showing strong seasonality, one should avoid random sampling from historical observation trajectories and replace it e.g.\ with random sampling from the same season as the forecasts to be reshuffled, or more sophisticated techniques such as mdSSh and simSSh.

In a nutshell, our overall findings indicate that there are generally only minor differences in the predictive performance of the various multivariate post-processing methods, and that the widely used ECC-Q approach constitutes a powerful benchmark method. Its straightforward applicability and very low computational costs make it a natural first choice to apply in future comparative studies of multivariate post-processing methods. Our findings are well in line with the results of the simulation studies performed in \citet{lbm20}, who also did not observe a single consistently best methods across all considered potential misspecifications. In the interest of ensuring the direct comparability of the methods, we restricted all approaches to the fixed sample size of the raw ensemble. That said, one advantage of the SSh variants is that they in principle allow for generating post-processed ensemble forecasts of arbitrary size, which might be advantageous for better modeling extreme events \citep{ltrg17} and offers a natural starting point for future research, for example by investigating the effect of the sample size on the performance in terms of recently proposed weighted multivariate proper scoring rules \citep{agz22}.

The case studies considered here provide several avenues for further generalization and analysis. While we have restricted our attention to temporal dependencies between lead times from 1 to 10 days, it would be interesting to also systematically compare the predictive performance in terms of spatial or inter-variable dependencies. Further, our focus was on copula-based two-step approaches to multivariate post-processing. Alternative methods based on parametric models for the full joint distribution \citep{bm15,frg19} or quantile mapping \citep{wzjz21} have been proposed and offer a natural starting point for further comparisons.

Recent research in post-processing has demonstrated the benefits of incorporating additional predictors on the forecasting performance of univariate methods, see e.g.\ \citet{rl18}. While these advanced post-processing methods can serve as building blocks of multivariate post-processing schemes, incorporating additional predictor information in the second, copula-based step is challenging, calling for the development of tailored approaches to machine learning methods for multivariate post-processing. There have been first studies in this direction focusing on obtaining spatially coherent forecast fields via generative adversarial networks \citep{dh21}.

Finally, the evaluation of multivariate predictive performance remains a challenging problem, and different multivariate evaluation metrics result in different rankings of the considered approaches. While there has been recent progress on the methodological aspects of multivariate evaluation \citep{zb19,ach22,agz22}, the need for systematic comparisons of the discrimination ability of multivariate proper scoring rules, ideally based on standardized benchmark datasets constitutes and important pathway towards a better understanding of advantages and disadvantages of individual metrics. As noted in \citet{lbm20}, post-processing studies based on large datasets such as the one investigated here, might provide helpful insights in this regard.

\bigskip
\noindent
{\bf Acknowledgments.} \  The authors gratefully acknowledge support by the Deutsche For\-schungsgemeinschaft (DFG) through project MO-3394/1-1 ``Statistische Nachbearbeitung von Ensemble-Vorhersagen f\"ur verschiedene Wettervariablen''. S\'andor Baran is further supported by the National Research, Development and Innovation Office under Grant No. NN125679. Sebastian Lerch gratefully acknowledges support by the Vector Stiftung through the Young Investigator Group ``Artificial Intelligence for Probabilistic Weather Forecasting''.

\end{document}